\documentclass[draftcls, 12pt,onecolumn,oneside,compress]{IEEEtran}
\usepackage{times,amsmath,color,amssymb,graphicx,epsfig,cite,psfrag,subfigure,algorithm,balance}
\usepackage{amsfonts,pifont,enumerate,cases}
\usepackage{mathrsfs} 
\usepackage[table]{xcolor} 
\usepackage{verbatim} 
\usepackage{geometry}
\newtheorem{property}{Property}

\newtheorem{remark}{\underline{Remark}}

\newtheorem{proof}{Proof}
\DeclareMathOperator*{\argmax}{argmax}

\newcommand{\tu}{\textup}

\begin{document}
\title{A Simple DFT-aided Spatial Basis Expansion Model and Channel Estimation Strategy for TDD/FDD Massive MIMO Systems}
\author{Hongxiang Xie, Feifei Gao, Shun Zhang, and Shi Jin
\thanks{H. Xie and F. Gao are with  Tsinghua National Laboratory for Information Science and Technology (TNList) Beijing 100084, P. R. China (e-mail: xiehx14@mails.tsinghua.edu.cn, feifeigao@ieee.org).}
\thanks{S. Zhang  is with the State Key Laboratory of Integrated Services Networks,
Xidian University, Xi'an 710071, P. R. China (email:
zhangshunsdu@gmail.com).}
\thanks{S. Jin is with the National Communications Research Laboratory, Southeast University, Nanjing 210096, P. R. China (email: jinshi@seu.edu.cn).}
}
\maketitle
\thispagestyle{empty}
\vspace{-20mm}
\begin{abstract}
This paper proposes a new transmission strategy for the multiuser
massive multiple-input multiple-output (MIMO) systems, including uplink/downlink
channel estimation and user scheduling for data transmission.
A discrete Fourier transform (DFT) aided spatial basis expansion model (SBEM) is
first introduced to represent the uplink/downlink channels with much few
parameter dimensions by exploiting  angle reciprocity and
the physical characteristics of the uniform linear array (ULA).
With SBEM, both uplink and downlink channel estimation of multiusers
can be carried out with very few amount  of training resources,
which significantly reduces the training overhead and feedback cost.
Meanwhile, the pilot contamination problem in the uplink training is
immediately relieved by exploiting the  spatial information of users.
To enhance the spectral efficiency and to fully utilize the spatial resources,
we also design a greedy user scheduling scheme during the data transmission period.
Compared to existing low-rank models, the newly proposed SBEM
offers an alternative for channel acquisition without need of channel statistics
for both TDD and FDD systems based on the angle reciprocity. Moreover, the proposed method can be efficiently deployed by
the fast Fourier transform (FFT). Various numerical results are  provided to corroborate the proposed studies.
\end{abstract}

\vspace{-5mm}

\section{Introduction}
Large-scale multiple-input multiple-output (MIMO) or ``massive MIMO'' systems \cite{marzetta2010noncooperative}
have drawn considerable interest from both academia and industry. Theoretically,
massive MIMO systems can almost perfectly relieve the inter-user interference in multiuser MIMO (MU-MIMO) systems with simple linear transceivers \cite{han(9)}.
It was shown in \cite{scalingupMIMO} that each antenna element of a very large MIMO system consumes exceedingly low power, and the total power can be made inversely proportional to the
number of antennas. Other advantages, such as high spectral efficiency,
security, robustness or reliable linkage, also play key roles in promoting massive MIMO systems more appealing  for
the next generation of wireless systems \cite{scalingupMIMO, MIMOnextgen}.

However, all these potential gains of massive MIMO systems rely heavily on the perfect
channel state information (CSI) at the base station (BS).
From the conventional orthogonal training  strategy \cite{han(13)}, the optimal number of training streams  should be the same as the number of the
transmit antennas and the length of the training should be no less than the number of
transmit antennas. Hence,  downlink training in massive MIMO system requires huge number of orthogonal training sequences.
This severe overhead as well as the accompanied high calculation complexity will overwhelm the system performance and mitigate any possible improvement.
For uplink training, the conventional channel estimation methods are generally feasible. However, as the number of users or the number of
user's antennas grows, the increased pilot overhead will
deteriorate the system efficiency and becomes the system bottleneck. If the non-orthogonal sequences are used for uplink training, then the so caused pilot contamination will also deteriorate the system performance.

For time division duplexing (TDD) massive MIMO systems,
downlink CSI can be obtained by leveraging the channel reciprocity \cite{jinshi[25]}, which has promoted quite many research works \cite{shift,EVD,Blind,CMT-based,summary6,data-aided}.
However, in practice the reciprocity between uplink and downlink may not exactly hold even for TDD systems due to the calibration error between the downlink/uplink RF chains \cite{challengetrend}.
In addition, the property of channel reciprocity has been proven to be robust only for the single-cell
scenario \cite{summary21}, and this will undoubtedly increase the pressure of multi-cell coordination.
On the other side, the downlink channel estimation for frequency division duplexing (FDD) massive MIMO system is always deemed as a difficult problem since the channel reciprocity does not hold and cannot be
used to simplify the estimation.
The authors of \cite{summary5,summary17,han(9)} applied the closed-loop training schemes
to sequentially design the optimal pilot beam patterns. The compressive sensing (CS) based feedback-reduction  in \cite{summary7}
and the distributed compressive channel estimation in \cite{KLT}  exploited the channel
statistics to reduce the heavy burden of feeding back large amount of measurements.

Except for the above regular attempts, a new way to design the transmission strategy for massive MIMO system
is to exploit the low-rank approximation of channel covariance matrix \cite{yin,Caire,jinshi}.
Based on the assumption that the angular spread (AS) of the
incident signals at BS from each user is narrow and the antenna elements only have
half-wave length spacing, the authors in \cite{yin,Caire,jinshi}  proposed to
reduce the dimensionality of the effective channels through eigen-decomposition of channel covariance matrices.
All these covariance-aware
methods could be categorized into \emph{spatial division multiplexing}
that utilizes non-overlapping  spatial information of different users
to realize the orthogonal transmission.
Meanwhile, compressed channel sensing
has been widely adopted for channel sparsity models, such as the Karhunen-Loeve transform basis representation in \cite{KLT}
and the virtual channel representation in \cite{CCS}. As the channel sparsity patterns
in these existing methods are often assumed unknown, high-complexity nonlinear CS reconstruction procedures are thus
inevitable.

In this paper, we propose an alternative low-rank model based on the discrete Fourier transform (DFT) of the steering
vectors for the uniform linear array (ULA). The new model could exploit the  spatial information
of the users  and is then named as spatial basis expansion model (SBEM).
It is shown that the uplink/downlink channel estimation of multiusers can be
carried out with very few training resources, and thus the overhead of training and feedback can also be reduced significantly.
Meanwhile, the pilot contamination in  uplink training can be immediately relieved. To enhance the spectral efficiency during
the data transmission, we propose a greedy user scheduling algorithm where users with orthogonal spatial information are allowed to transmit simultaneously. 
Compared to the existing low-rank models \cite{Caire,KLT,CCS,yin}, the proposed one offers following several benefits:
(1) SBEM can exploit the angle reciprocity and simplify the DL training from the UL channel estimation
  for both TDD/FDD massive MIMO systems;
(2) SBEM does not need the knowledge of channel covariance;
(3) SBEM could be simply implemented by the fast Fourier transform (FFT) and the partial FFT \cite{partialFFT1}.
By contrast, \cite{yin} and \cite{Caire} need EVD for high-dimensional covariance matrices, while \cite{KLT}
and \cite{CCS} require nonlinear optimization.
Various numerical results are provided to evaluate performances of the proposed method.

%
%
%

The rest of the paper is organized as follows. In section \ref{sec:modelandproperty}, the system model as well as the channel characteristics with narrow incident
signals are described.
The SBEM aided channel estimations for uplink/downlink transmission are presented in  section \ref{sec:Uplinktraining}. Section \ref{sec:datatransmission} designs  the user scheduling for data transmission, followed by
simulations in section \ref{sec:simulation}. Finally, conclusions are drawn in section \ref{sec:conclusions}.

\textbf{Notations:} Vectors and matrices are boldface small and
capital letters;  the transpose, complex conjugate, Hermitian,
inverse, and pseudo-inverse of the matrix ${\mathbf A}$ are denoted
by ${\mathbf A}^T$, ${\mathbf A}^*$, ${\mathbf A}^H$, ${\mathbf
A}^{-1}$ and ${\mathbf A}^\dag$, respectively; ${\rm tr}({\mathbf
A})$ is the trace of ${\mathbf A}$; $[{\mathbf A}]_{ij}$ is the $(i,j)$th entry of
${\mathbf A}$;  the entry index of vector and matrix starts from $0$;  diag$\{{\mathbf a}\}$ denotes a diagonal matrix
with the diagonal elements constructed from ${\mathbf a}$, while  diag$\{\mathbf A\}$
denotes a vector whose elements are extracted from the diagonal components of $\mathbf A$;
${\mathbf I}$ is the
identity matrix with appropriate size, and $\mathbb{E}\{\cdot\}$ is the statistical
expectation; $\lceil x\rceil$ denotes the
smallest integer no less than $x$, $\lfloor x\rfloor$ denotes the
largest integer no more than $x$ and $\lfloor x\rceil$ denotes the integer closest to $x$;
$\left[\mathbf H\right]_{:,\mathcal{D}}$ denotes the
sub-matrix of $\mathbf H$ by collecting the columns indexed by $\mathcal{D}$,
and $\left[\mathbf H\right]_{\mathcal{D},:}$ denotes the
sub-matrix of $\mathbf H$ by collecting the rows indexed by $\mathcal{D}$;
$\left[\mathbf h\right]_{\mathcal{D},:}$ indicates
the sub-vector of $\mathbf h$ by keeping the elements indexed by $\mathcal{D}$;
$|\mathcal{D}|$ denotes the cardinality of the set
$\mathcal{D}$; and
$\left\|\mathbf h\right\|$ denotes the Euclidean norm of $\mathbf h$;
``$\backslash$" defines the set subtraction operation.

\section{System Model and Channel Characteristics}\label{sec:modelandproperty}
\subsection{System Model}\label{sec:system model}
Array structure based physical channel models have been widely adopted for MIMO systems,
such as the spatial channel model (SCM) \cite{SCM}, which exploits the array manifold and information of the
direction of arrivals (DOA) as well as the direction of departures (DOD) of propagation signals.
For massive MIMO systems, the significantly improved spatial resolution of large-scale antenna arrays as
well as plenty of emerging low-complexity DOA acquisition techniques \cite{liulingjia0,liulingjia1,liulingjia2,2D-ESPRIT}
have further promoted these physical channel models \cite{Caire,yin,jinshi}.
Many  works \cite{challengetrend,jinshi,five} of massive MIMO systems
require BS to be elevated at a very high altitude, say on the top of a high building or a dedicated tower such that
there are few surrounding scatterers at the end of BS. In this case, the incident angular spread seen by the BS array is usually limited in a narrow region \cite{yin,jinshi,Caire}.

Let us consider a multiuser massive MIMO system, where BS is equipped with $M\gg 1$
antennas in the form of ULA,
and $K$ single-antenna users are
randomly distributed in the coverage area. For better illustration of our key idea,
we assume the channel is  flat fading for the time being
as did in \cite{yin,Caire}.

\begin{figure}[t]
      \centering
     \includegraphics[width=100mm]{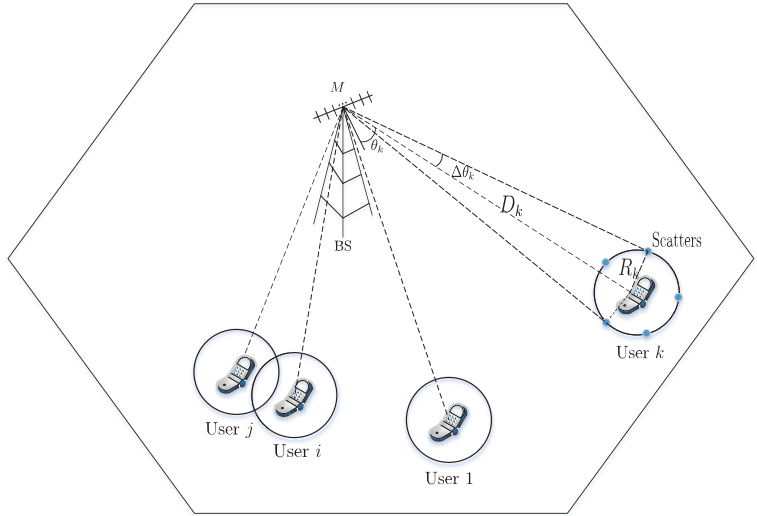}
     \caption{System of one-ring model. Users are randomly distributed and surrounded by $P$ local scatterers.
     The mean DOA and AS of user-$k$ are $\theta_k$ and $\Delta\theta_k$, respectively.}
     \label{fig:onering}
\end{figure}

Consider the classical ``one-ring model'' 
adopted in \cite{Caire},  where  user-$k$ located at $D_k$ meters away from BS is surrounded by a ring of $P\gg 1$ local scatterers (see Fig. \ref{fig:onering}) with the radius $R_k$.
Then the propagation from user-$k$ to BS is composed of $P$ rays and the corresponding $M\times 1$ uplink channel can be expressed as \cite{yin}:
\begin{equation}\label{equ:channelmodel}
\mathbf h_{k}=\frac{1}{\sqrt{P}}\sum_{p=1}^P \alpha_{kp}\mathbf a(\theta_{kp}),
\end{equation}
where $\alpha_{kp}\sim\mathcal{CN}(0,\xi_{kp})$ represents the complex gain
of the $p$-th ray and is independent and identically distributed (i.i.d.) from each other. Moreover, $\mathbf a(\theta_{kp})\in\mathbb{C}^{M\times 1}$ is the array manifold vector and has the form
\begin{equation}\label{equ:steeringvector}
\mathbf a(\theta_{kp})=\left[1,e^{j\frac{2\pi d}{\lambda}\sin\theta_{kp}},\ldots,e^{j\frac{2\pi d}{\lambda}(M-1)\sin\theta_{kp}}\right]^T,
\end{equation}
where $d$ is the antenna spacing, $\lambda$ denotes the signal carrier wavelength, and $\theta_{kp}$ represents
the DOA of the $p$-th ray.

Since the distance $D_k$ from BS to user-$k$ is always much larger than the radius $R_k$, the incident rays
will be constrained within a narrow angular spread (AS) $\Delta\theta_k\approx\arctan(R_k/D_k)$ \cite{Caire}. In other words, the incident angles of user-$k$  with mean DOA
$\theta_k$ is limited in the narrow angular range $[\theta_k-\Delta\theta_k,\theta_k+\Delta\theta_k]$.
Within this narrow angular range, it is obvious that $\mathbf{a}(\theta_{kp})$'s are highly correlated with each other such that the channel covariance matrix of user-$k$,
defined as $\mathbf{R}_{h_k}=\text{E}\{\mathbf{h}_k\mathbf{h}_k^H\}$ approximately possesses the low-rank property. Based on this low-rank property, \cite{KLT}, \cite{yin},  \cite{Caire}
and \cite{CCS} presented low-rank channel models and designed the corresponding channel estimation to reduce the channel overhead as well as computational complexity. However, the methods in\cite{yin} and \cite{Caire} need eigen-decomposition for an $M\times M$ channel covariance matrix, while \cite{KLT}
and \cite{CCS} demand for nonlinear optimization, making their methods still complex for practical implementations.

In this paper, we present an alternative way to exploit the such a low-rank property, targeting at further reducing the channel estimation complexity. Due to the high correlations among  $\mathbf{a}(\theta_{kp})$, $p=1,\ldots, P$, $\mathbf{h}_k$ can  be expanded from some orthogonal basis as
\begin{equation}\label{equ:BEM}
\mathbf{h}_k=\sum_{q=1}^\nu \kappa_q \mathbf{b}_q, \ \textup{for}\  k=1,\ldots,K,
\end{equation}
where $\mathbf{b}_q$'s are basis vectors to be determined and $\kappa_q$'s are the corresponding coefficients.
Equation (\ref{equ:BEM}) is also known as basis expansion model (BEM)  as long as
we could find a  set of uniform basis vectors $\mathbf{b}_q$'s for any possible  $\mathbf{h}_k$ \cite{temporalBEM1}.
Then the task of channel estimation will be greatly simplified and will be converted to estimating $\nu$ constants $\kappa_q$'s only.

\subsection{Characteristics of ULA and Channel Vectors}\label{sec:ULAcharacteristics}
Define the normalized DFT of the channel vector $\mathbf{h}_k$
as $\tilde{\mathbf h}_k=\mathbf{F}\mathbf{h}_k$, where $\mathbf F$ is the $M\times M$ DFT matrix
whose $(p,q)$th element is  $\left[\mathbf F\right]_{pq}=e^{-j\frac{2\pi}{M}pq}/\sqrt{M}$.

\begin{property} \label{property:1}
    For the simplest 1-ray case, i.e., $\mathbf h_k=\alpha_k\mathbf a(\theta_k)$, $\tilde{\mathbf h}_k$
    is approximately a sparse vector that contains the  spatial information of the impinging signal.
\end{property}

\begin{figure}[t]
\centering
\subfigure[$\tilde{\mathbf h}_k$ of single incident ray with $\theta_k=37.54^\circ$]{
\includegraphics[width=0.45\textwidth]{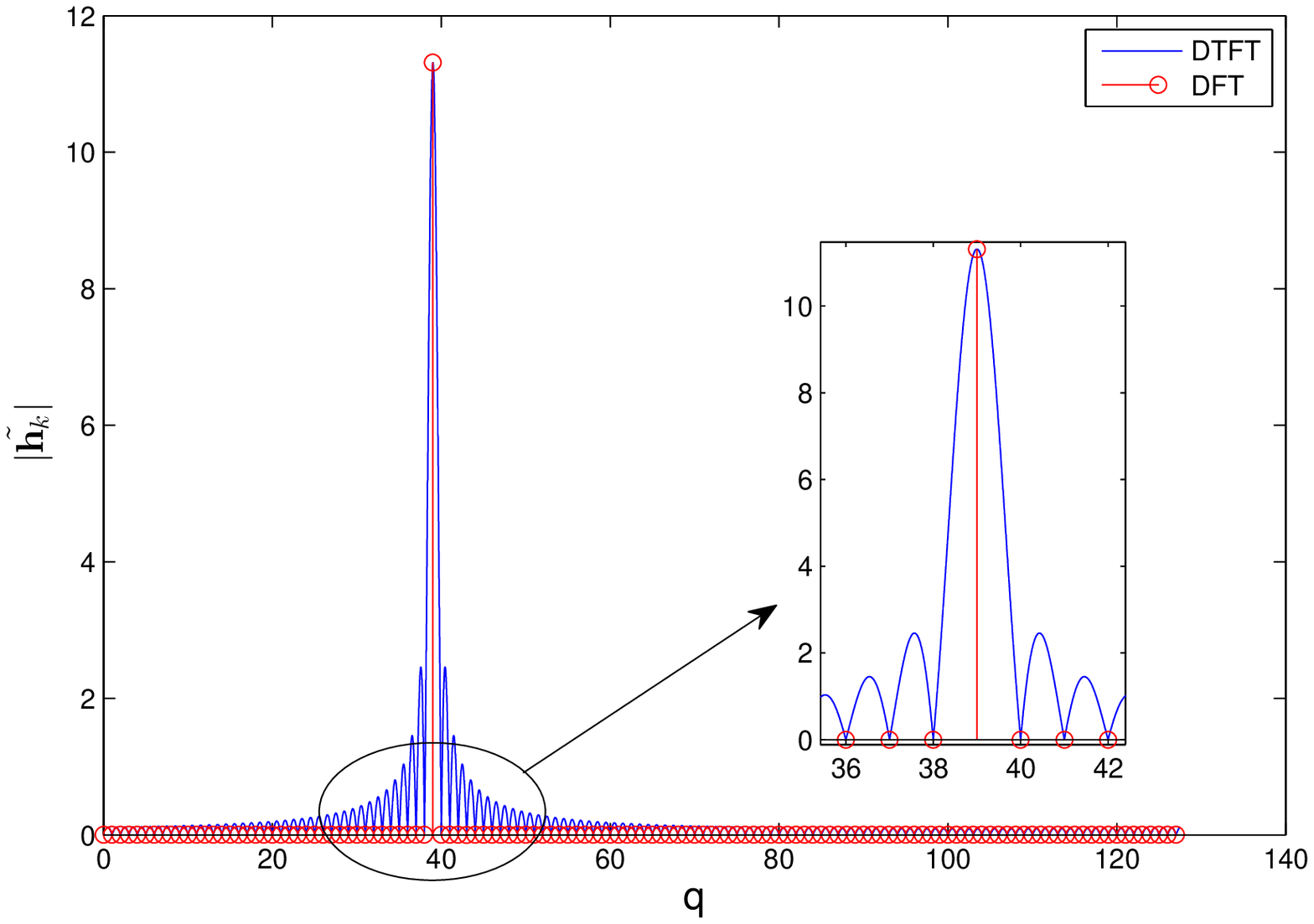}
\label{fig:2a}
}
\subfigure[$\tilde{\mathbf h}_k$ of single incident ray with $\theta_k=37^\circ$]{
\includegraphics[width=0.45\textwidth]{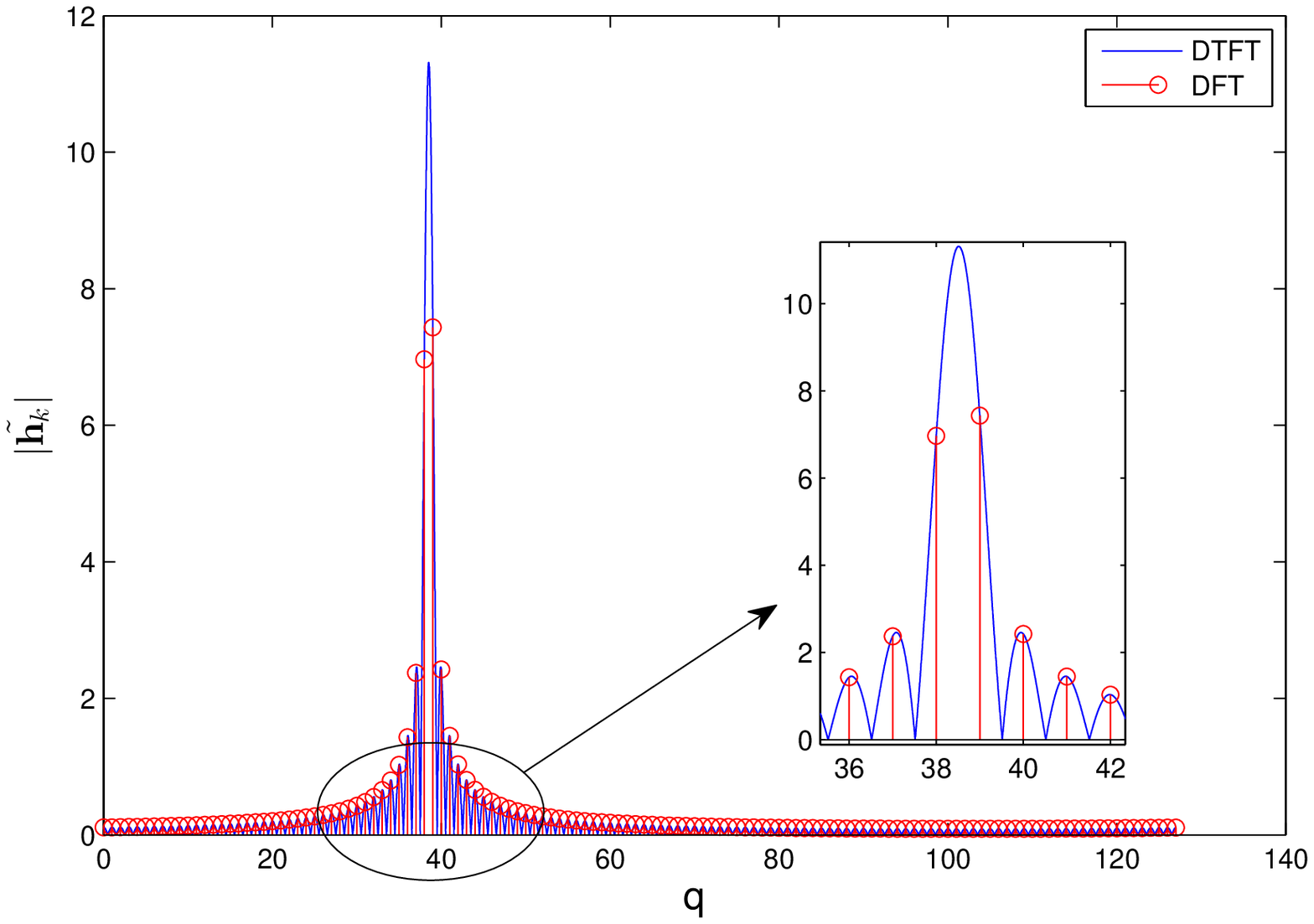}
\label{fig:2b}
}
\caption{Examples of $\tilde{\mathbf h}_k$ for 1-ray case, where $d=\frac{\lambda}{2}$ and $M=128$.}
\label{fig:singlemultipath}
\end{figure}

\begin{proof}\label{proof:p1}
 The $q$-th component of $\tilde{\mathbf h}_k$ is computed as
\begin{equation}\label{equ:channeldftbjk}
\begin{aligned}
\left[\tilde{\mathbf h}_k \right]_{q}&=
\frac{\alpha_k}{\sqrt{M}}\sum_{m=0}^{M-1}e^{-j\left(\frac{2\pi}{M}mq-\frac{2\pi}{\lambda}m d \sin\theta_{k}\right)}\\
&=\frac{\alpha_k}{\sqrt{M}}e^{-j\frac{M-1}{2}\left[\frac{2\pi}{M}q-\frac{2\pi}{\lambda}d\sin\theta_{k}\right]}\cdot
\frac{\sin\left[\left(\frac{2\pi}{M}q-\frac{2\pi}{\lambda}d\sin\theta_{k}\right)\frac{M}{2}\right]}
{\sin\left[\left(\frac{2\pi}{M}q-\frac{2\pi}{\lambda}d\sin\theta_{k}\right)\frac{1}{2}\right]}.
\end{aligned}
\end{equation}
When $M\frac{d}{\lambda}\sin\theta_k$ equals to some integer $q_0$, then $\tilde{\mathbf h}$ only has one non-zero element $[\tilde{\mathbf h}_k]_{q_0}=\alpha_k\sqrt{M}$. In this specific case, $\tilde{\mathbf h}_k$ is highly sparse and all powers are concentrated on the $q_0$-th DFT point, as illustrated
in Fig. \ref{fig:2a}. However, for most other cases, $M\frac{d}{\lambda}\sin\theta_k$ is not an integer,  and the channel power will leak from  the $(\lfloor M\frac{d}{\lambda}\sin\theta_k\rceil)$-th DFT point to other DFT points, as shown in Fig. \ref{fig:2b}. In fact, DFT outputs
are discrete samples of discrete-time Fourier transform (DTFT) of $\mathbf a(\theta_k)$, i.e., a Sinc function,  at the points of $\frac{2\pi q}{M}$, $q=0,\ldots,M-1$.
It is then  easily
known that the degree of leakage is inversely proportional to $M$ but is proportional to the deviation $ \left(M\frac{d}{\lambda}\sin\theta_k-\lfloor M\frac{d}{\lambda}\sin\theta_k\rceil\right)$.
Hence when $M$ is sufficiently large,  $\tilde{\mathbf h}_k$ can still be approximated by a sparse vector with most of power being concentrated around $M\frac{d}{\lambda}\sin\theta_k$.  Specifically, for the ideal case $M\rightarrow \infty$, there always exists a $q_0$ that satisfies $q_0=M\frac{d}{\lambda}\sin\theta_k$ for any possible $\theta_k$ and then the power leakage is eliminated.
\end{proof}

From \emph{Property \ref{property:1}}, it is clear that
most channel power is concentrated on DFT points in the vicinity of $\lfloor M\frac{d}{\lambda}\sin\theta_k\rceil$ when $M$ is large. It is then of interest
to find the minimal number of DFT points around  $\lfloor M\frac{d}{\lambda}\sin\theta_k\rceil $ that contain  at least $\eta$ percentage of the total power.
Define the corresponding set of DFT points as $\mathcal{C}_k$, which  can be found from the following optimization:
\begin{equation}\label{equ:powerratio}
\begin{aligned}
 \min_{\mathcal{C}_k}&\ \ \ \ C_k\triangleq|\mathcal{C}_k| \\
  \tu{s.t.}\ &\ \ \sum_{q\in \mathcal{C}_k}\frac{1}{M}\left\{
  \frac{\sin\left[\left(\frac{2\pi}{M}q-\frac{2\pi}{\lambda}d\sin\theta_{k}\right)\frac{M}{2}\right]}
{\sin\left[\left(\frac{2\pi}{M}q-\frac{2\pi}{\lambda}d\sin\theta_{k}\right)\frac{1}{2}\right]}
\right\}^2\geq \eta.
\end{aligned}
\end{equation}
The optimization in \eqref{equ:powerratio} is difficult to solve in closed-form. Nevertheless, since \eqref{equ:powerratio}
is merely dependent on the array configuration, we can
establish the off-line tables for  $\theta_k\in(-\frac{\pi}{2},\frac{\pi}{2})$ and some pre-defined $\eta$'s.  An example of the off-line table of $C_k$ for a single incident ray is given
in Tab. \ref{table:offlinetable} with $M=128$ and $\eta=95\%$,  where $C_k$ is a function $\theta_k$ and $\eta$.
It is seen from Tab.~\ref{table:offlinetable} that the maximal number of leakage points that contain at least $95$\% of the channel
power is $10$ for any incident angle, which is very small compared to the total antenna number $M=128$. Hence, we could safely treat $\tilde{\mathbf{h}}_k$ as a sparse vector for massive antenna system.

\begin{table}
\centering
\caption{An example of off-line table of $C_k$, where $M=128$, $\eta=0.95$ and $\theta_k\in[1^\circ,\ 90^\circ]$}\label{table:offlinetable}
\rowcolors{1}{white}{gray!25}
\begin{tabular}{|c|c|c|c|c|c|c|c|c|c|c|c|c|c|c|c|c|c|}
\hline
$\theta_k$   &  1$^\circ$    &  3$^\circ$    &   5$^\circ$     &   7$^\circ$     &   9$^\circ$     &  11$^\circ$    &  13$^\circ$   & 15$^\circ$
             & 17$^\circ$    &  19$^\circ$   &  21$^\circ$     &   23$^\circ$    &   25$^\circ$    &   27$^\circ$   &  29$^\circ$ \\
\hline
 $C_k$  &      1    &  8    &   9   &   4   &   1 &   5   &  9  &   9    &   7  &   4   &  1    &   1   &   1    &   1  &   1 \\
\hline \hline
$\theta_k$    &  31$^\circ$     &   33$^\circ$   &   35$^\circ$ &  37$^\circ$     &  39$^\circ$     &   41$^\circ$    &   43$^\circ$  & 45$^\circ$ &  47$^\circ$     &  49$^\circ$    &   51$^\circ$  &   53$^\circ$ &  55$^\circ$     &  57$^\circ$      &   59$^\circ$  \\
\hline
$C_k$    &   1    &   3    &   7  &   10    &  6    &  1    &   8   &   6  &   4   &   7    &   6   &   1 &   9  &  7    &   3  \\
  \hline\hline
 $\theta_k$   & 61$^\circ$     &   63$^\circ$     &  65$^\circ$    &  67$^\circ$    &   69$^\circ$  &   71$^\circ$  &  73$^\circ$   &  75$^\circ$
              & 77$^\circ$     &   79$^\circ$     &   81$^\circ$   &  83$^\circ$    &  85$^\circ$   &   87$^\circ$  &   89$^\circ$ \\
  \hline
   $C_k$  &   1  &  1  &   1  &   1    &  6   &   10 &   4   &  4   &   8    &   4   &   5  &   9   &   5   &   1    &   1\\
 \hline
\end{tabular}
\end{table}

\begin{property} \label{property:2}
For the multi-ray case \eqref{equ:channelmodel},  let us  define $\mathcal{D}_k$ as the index set of the continuous DFT points that contain at least $\eta$ percentage of the total channel power. The upper bound of the cardinality of $\mathcal{D}_k$ can be expressed as
    \begin{equation}\label{equ:deltaB}
      |\mathcal{D}_k|\leq \lceil2M\frac{d}{\lambda}\cdot|\cos\theta_k|\cdot\Delta\theta_k+1\rceil+
      C_{\max},
    \end{equation}
    where  $C_{\max}\triangleq \max\  C_k$ for given $\eta$ and all $ \theta\in[\theta_k-\Delta\theta_k,\theta_k+\Delta\theta_k]$.
\end{property}
\begin{proof}\label{proof:p2}
The left bound of $\mathcal{D}_k$ is determined by the DFT of the leftmost ray with $\theta_{kp}=\theta_k-\Delta\theta_k$ and
can be expressed as $\lfloor M\frac{d}{\lambda}\sin(\theta_k-\Delta\theta_k)\rfloor-\lceil C_{\max}/2\rceil $ according to
\emph{Property~\ref{property:1}}.
Similarly, the right bound of $\mathcal{D}_k$ depends on the DFT of the rightmost ray with $\theta_{kp}=\theta_k+\Delta\theta_k$ and can be expressed as
$\lceil M\frac{d}{\lambda}\sin(\theta_k+\Delta\theta_k)\rceil+\lceil C_{\max}/2 \rceil$.
Then for any single ray with incident DOA inside $[\theta_k-\Delta\theta_k,\theta_k+\Delta\theta_k]$,
the corresponding main leakage points that count for  $\eta$ percentage of its own power will be included in $\mathcal{D}_k$. 
From the above discussion, the upper bound of the cardinality of $\mathcal{D}_k$ can be directly obtained as
\begin{equation}\label{equ:multi-rayrange}
\begin{aligned}
       |\mathcal{D}_k|&\leq \lceil M\frac{d}{\lambda}\sin(\theta_k+\Delta\theta_k) \rceil-
       \lfloor M\frac{d}{\lambda}\sin(\theta_k-\Delta\theta_k) \rfloor +1+C_{\max} \\
       &\leq \lceil2M\frac{d}{\lambda}\cdot|\cos\theta_k|\cdot\Delta\theta_k+1\rceil+C_{\max}.
\end{aligned}
\end{equation}
\begin{figure}
\centering
\includegraphics[width=100mm]{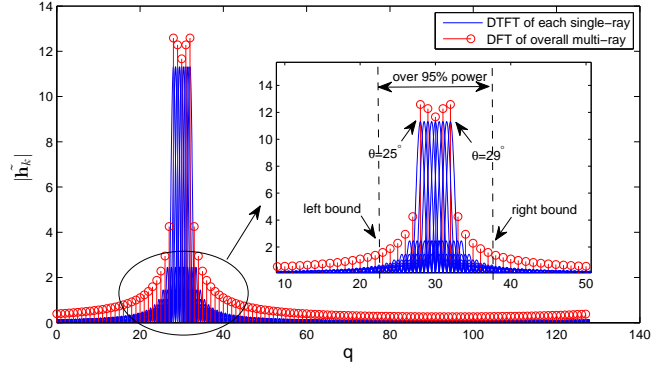}
\caption{Example of multi-ray channel with incident angles in $[25^\circ,29^\circ]$. DTFT of each single-ray channel (blue lines) as
well as DFT of the overall multi-ray channel (red points) are depicted, respectively.
\label{fig:multi-raypower}}
\end{figure}
\end{proof}

An example of a $9$-ray channel with incident angles in $[25^\circ,29^\circ]$ is given in Fig. \ref{fig:multi-raypower},
where the  DTFT of each single ray   as well as the DFT of the overall multi-rays
are depicted respectively. The numerical results in this example show that  the actual cardinality of $\mathcal{D}_k$ containing $95\%$ power is $15$, and the upper
bound calculated from \eqref{equ:multi-rayrange} is also $15$.

Following \emph{Property \ref{property:2}} and bearing in mind that $\Delta\theta_k$ is small, it is readily known  that
$|\mathcal{D}_k|\leq \left\lceil\frac{2 Md}{\lambda}\cdot|\cos\theta_k|\cdot\Delta\theta_k+1\right\rceil+ C_{\max}$
is still small compared to $M$. Hence, $\tilde{\mathbf h}_k$ corresponding to the multi-ray case is also approximately sparse with most power being  contained in limited number of entries.

\begin{remark}
 The sparsity of mutli-ray case is very obviously if $M\rightarrow\infty$ is assumed, as did in most massive MIMO works \cite{yin,jinshi}. Nevertheless, we do not make such an ideal assumption throughout this paper in order to present a practical solution for large but limited number of array antennas.
\end{remark}

Therefore, the key idea of this paper that is to approximate the channel vector with fewer parameters as
\begin{equation}\label{equ:preambleapprox}
\begin{aligned}
\mathbf h_{k}&=\mathbf F^H\tilde{\mathbf h}_{k}
  \approx\left[\mathbf F^H\right]_{:,\mathcal{D}_{k}}\left[\tilde{\mathbf h}_{k}\right]_{\mathcal{D}_{k},:}
    =\sum_{q\in\mathcal{D}_k}\tilde{h}_{k,q}\mathbf f_q,
\end{aligned}
\end{equation}
where $\tilde{h}_{k,q}\triangleq [\tilde{\mathbf h}_k]_q$ denotes the $q$-th element of $\tilde{\mathbf h}_k$ while $\mathbf f_q$
is the $q$-th column of $\mathbf F^H$.
Comparing  with \eqref{equ:BEM}, the expansion in  \eqref{equ:preambleapprox} is also in the form of BEM where the basis vectors $\mathbf{b}_q\triangleq\mathbf{f}_q$ are orthogonal to each other. Hence,  we only need to estimate the limited BEM parameters $\tilde{h}_{k,q}$.

Interestingly, the DFT vector $\mathbf f_q$ coincides with the steering vector as
$\mathbf f_q=\mathbf a(\theta_q)$ where $\theta_q=\arcsin\frac{q\lambda}{Md}$, which means that
$\mathbf f_q$ formulates an array beam towards
the physical direction $\theta_q=\arcsin\frac{q\lambda}{Md}$. Hence, all
 beams $\mathbf f_q,\ q\in\mathcal{D}_k$ will point towards the AS of user-$k$ and are orthogonal to each other. Consequently,
the beam indices $\mathcal{D}_k$ can be viewed as the \emph{spatial signature} of user-$k$ \cite{spatialsignature}, and \eqref{equ:preambleapprox} can be deemed as the
\emph{spatial BEM} (SBEM), as named after the popular temporal BEM \cite{temporalBEM1}.

\begin{property}\label{property:rotation}
Define
\begin{equation}\label{equ:rotatematrix}
 \mathbf \Phi(\phi)=\textup{diag}\left\{\left[1,e^{j\phi},\cdots,e^{j(M-1)\phi}\right]\right\},
\end{equation}
where $\phi\in[-\frac{\pi}{M},\frac{\pi}{M}]$ is a  \emph{shift parameter}. 
The  operation $\tilde{\mathbf h}_{k}^{\tu{ro}}=\mathbf F\mathbf \Phi(\phi)\mathbf h_k$ can further concentrate  channel power within fewer entries of $\tilde{\mathbf h}_{k}^{\tu{ro}}$ for certain value of $\phi$, and this operation is named as \emph{spatial rotation}.
\end{property}

\begin{figure}[t]
\centering
\subfigure[$\tilde{\mathbf h}_k$ with $|\mathcal{D}_k|=19$]{
\includegraphics[width=0.45\textwidth]{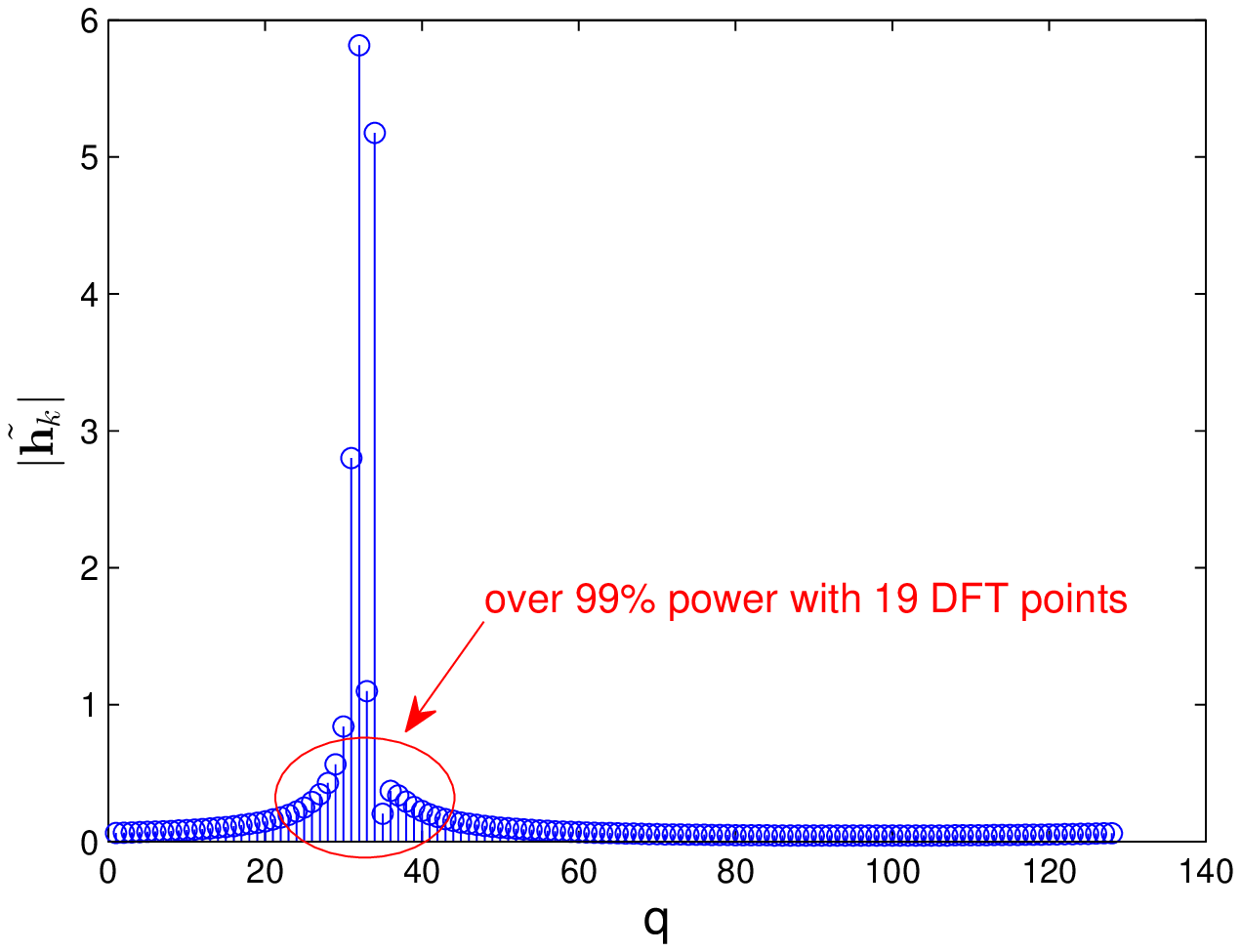}
\label{fig:3a}
}
\subfigure[$\tilde{\mathbf h}_k^{\tu{ro}}$ with $|\mathcal{D}_k^{\tu{ro}}|=6$]{
\includegraphics[width=0.45\textwidth]{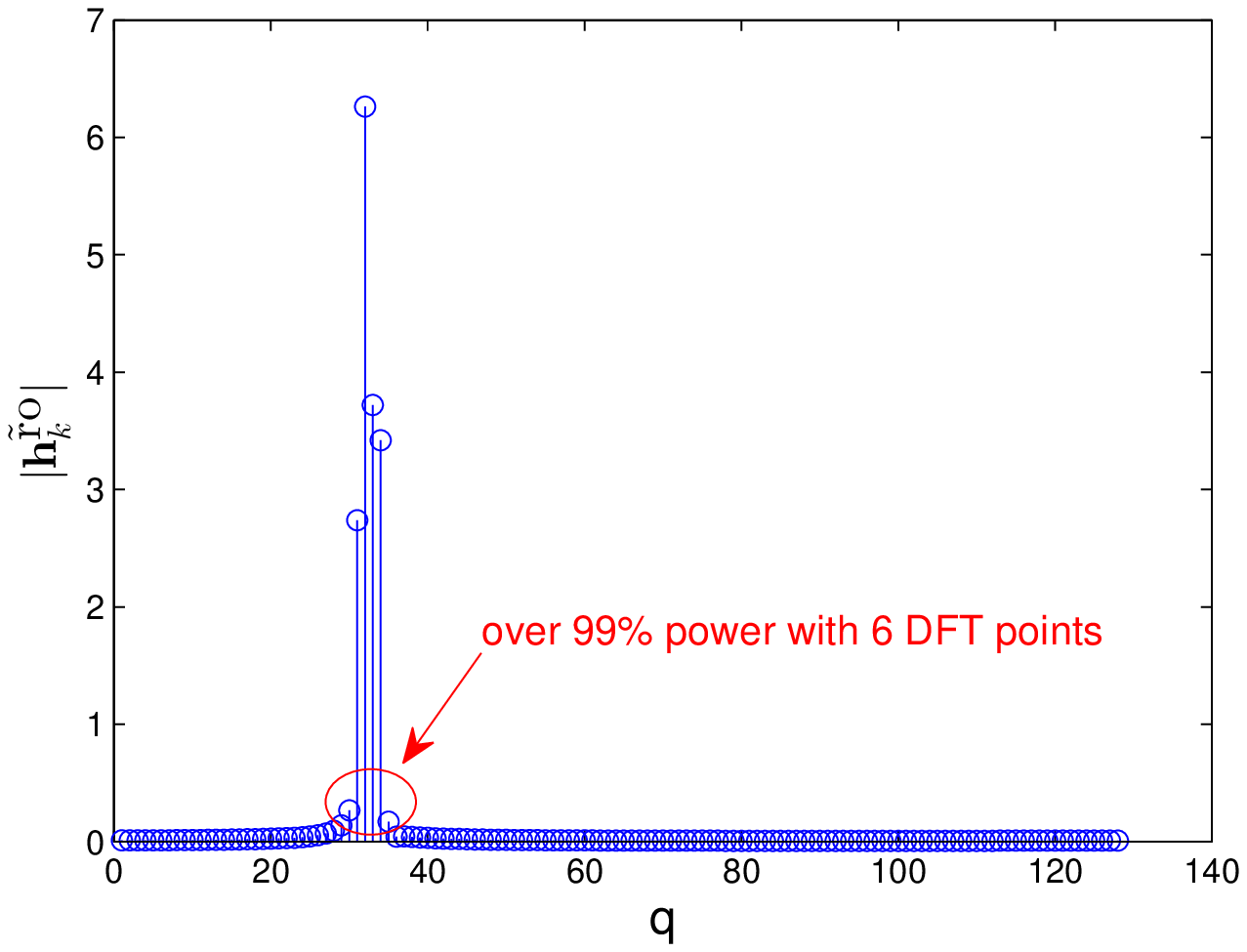}
\label{fig:3b}
}
\caption{Comparison of multi-ray channels with/without spatial rotation, where $\theta_k=30^\circ$, $\Delta\theta_k=2^\circ$,
$M=128,\ d=\frac{\lambda}{2}$, $\eta=0.99$
and the optimal $\phi_k=-0.0142$ (in $\tu{radian}$), within $[-\frac{\pi}{128},\frac{\pi}{128}]$. }
\label{fig:multipath}
\end{figure}

\begin{proof}
Consider the 1-ray case for the ease of illustration.
In Fig. \ref{fig:2b}, when the DOA of the incident signal is not $\arcsin\frac{q_0\lambda}{Md}$
for some integer $q_0$, i.e., mismatched with DFT points,
then power leakage will happen. Formulate a new channel vector as $\tilde{\mathbf h}_{k}^{\tu{ro}}=\mathbf F\mathbf \Phi(\phi)\mathbf h_k$.
According to the analysis in \emph{Property \ref{property:1}}, let us gradually change $\phi$  from $-\frac{\pi}{M}$ to $\frac{\pi}{M}$.\footnote{The reason for
$\phi_k\in[-\frac{\pi}{M},\frac{\pi}{M}]$ lies in that the resolution of DFT is $\frac{2\pi}{M}$.}
Then, there will exist a $\phi_k=(\frac{2\pi q_0}{M}-\frac{2\pi d}{\lambda}\sin\theta_k)$
that makes $\tilde{\mathbf h}_k^{\tu{ro}}$ possess only one non-zero element
$\tilde{h}_{k,q_0}^{\tu{ro}}$ at $q_0$.
For example in Fig. \ref{fig:singlemultipath}, spatial rotation with
$\phi_k=0.024$ radian can help to strengthen the sparsity of $\tilde{\mathbf h}_k$
in Fig.~\ref{fig:2b} to the form in Fig. \ref{fig:2a}.

For the multi-ray cases, we can also formulate $\tilde{\mathbf h}_{k}^{\tu{ro}}=\mathbf F\mathbf \Phi(\phi)\mathbf h_k$,
and define  $\mathcal{D}_k^{\tu{ro}}$ as the continuous index set such that
$[\tilde{\mathbf h}_{k}^{\tu{ro}}]_{\mathcal{D}_k^{\tu{ro}},:}$ contains at least $\eta$ percentage of the channel power.
Next, we can search $\phi$  from $-\frac{\pi}{M}$ to $\frac{\pi}{M}$
and select the optimal $\phi_k$ that minimizes $|\mathcal{D}_k^{\tu{ro}}|$. An example of multi-ray channel with $\theta_k=30^\circ$ and $\Delta\theta_k=2^\circ$ is given in Fig. \ref{fig:multipath},
where $M=128,\ d=\frac{\lambda}{2}$ and $\eta=0.99$.
It can be seen that the cardinality  of $\mathcal{D}_k^{\tu{ro}}$ is  only $|\mathcal{D}_k^{\tu{ro}}|=6$ after spatial rotation, while the
cardinality  before the rotation is  $|\mathcal{D}_k|=19$.
\end{proof}


 Based on the \emph{Property \ref{property:rotation}} and similar to \eqref{equ:preambleapprox}, there is
\begin{equation}\label{equ:refinedBEM}
  \mathbf h_{k}=\mathbf \Phi(\phi_k)^H\mathbf F^H\tilde{\mathbf h}^{\tu{ro}}_{k}
  \approx\mathbf \Phi(\phi_k)^H\left[\mathbf F^H\right]_{:,\mathcal{D}^{\tu{ro}}_{k}}\left[\tilde{\mathbf h}^{\tu{ro}}_{k}\right]_{\mathcal{D}^{\tu{ro}}_{k},:}
    =\sum_{q\in\mathcal{D}^{\tu{ro}}_k}\tilde{h}^{\tu{ro}}_{k,q}\mathbf \Phi(\phi_k)^H\mathbf f_q,
\end{equation}
where $\mathbf \Phi(\phi_k)^H\mathbf f_q,\ q\in\mathcal{D}_k^{\tu{ro}}$ are the new equivalent basis vectors.
Interestingly, these new basis vectors are still
mutually orthogonal, and thus \eqref{equ:refinedBEM} is also a type of SBEM.
In many cases, $|\mathcal{D}_k^{\tu{ro}}|$ is much less than $|\mathcal{D}_k|$ after the optimal rotation, and hence the number of
the channel parameters to be estimated is further reduced.

\begin{remark}
In fact, the operation $\mathbf \Phi(\phi_k)\mathbf h_k$ can be viewed as rotating the orthogonal beams $\mathbf f_q$'s by the same
angle such that the new beams $\mathbf \Phi(\phi_k)^H\mathbf f_q$'s
point towards user-$k$ more accurately while still keeping their  orthogonality.
\end{remark}

\begin{property}\label{remark:nonoptimalphase}
Define a set $\mathcal{E}_k^{\tu{ro}}$ that is sufficiently far from $\mathcal{D}_k^{\tu{ro}}$. Then, $[\mathbf F\mathbf{\Phi}(\phi)\mathbf h_k]_{\mathcal{E}_k^{\tu{ro}},:}$
always has ignorable value for any $\phi\in[-\pi/M,\pi/M]$ as compared with $[\mathbf F\mathbf{\Phi}(\phi)\mathbf h_k]_{\mathcal{D}_k^{\tu{ro}},:}$.  The reason is that $\mathbf F\mathbf{\Phi}(\phi)\mathbf h_k$ is the sampling of the DTFT
of $\mathbf h_k$ after being shifted by a small value $\phi$. Hence,  as shown in Fig.~\ref{fig:multipath}, $[\mathbf F\mathbf{\Phi}(\phi)\mathbf h_k]_{\mathcal{E}_k^{\tu{ro}},:}$
still possess very small value compared to the sampling points in ${\mathcal{D}_k^{\tu{ro}}}$.
This property will be used later during the channel estimation.
\end{property}

Nevertheless, the practical standards normally regulate a fixed number of channel parameters to be estimated instead of considering a dynamic number $|\mathcal{D}_k^{\tu{ro}}|$.
Denote $\tau$ 
 as the number of the channel parameters that the system  could handle, and define the  set containing continuous $\tau$ integers as $\mathcal{B}_k^{\tu{ro}}$. Then we should select $\mathcal{B}_k^{\tu{ro}}$ as well
as the shift parameter $\phi_k$ such that $\left[\tilde{\mathbf h}_k^{\tu{ro}}\right]_{\mathcal{B}^{\tu{ro}}_k,:}$ possesses the maximum channel power, i.e.,
\begin{align}\label{equ:phaserotationobjective}
  \max_{\phi_k,\ \mathcal{B}^{\tu{ro}}_k}\ \ &\left\|\left[\tilde{\mathbf h}_k^{\tu{ro}}\right]_{\mathcal{B}^{\tu{ro}}_k,:}\right\|^2
  \ \ \ \ \tu{subject to} \  |\mathcal{B}^{\tu{ro}}_k|=\tau.
\end{align}
The above optimization can be achieved simply by sliding window of size $\tau$
over the elements in $\tilde{\mathbf h}_k^{\tu{ro}}$ together by  a one dimensional search over $\phi\in[-\pi/M,\pi/M]$.

\section{Channel Estimation With SBEM }\label{sec:Uplinktraining}

Assume the current cell is allocated $\tau<K$ orthogonal training sequences of length $L<T$, where $T$ is the channel
coherence interval, for both uplink and downlink training. Denote the corresponding orthogonal training set in the considered cell as $\mathcal{S}_{\textup{cell}}=\{\mathbf s_1,\ldots,\mathbf s_{\tau}\}$ with
$\mathbf s_i^H\mathbf s_j=L\sigma_p^2\cdot\delta(i-j)$, where $\sigma_p^2$ is the pilot signal training  power.

\begin{figure}[t]
      \centering
     \includegraphics[width=100mm]{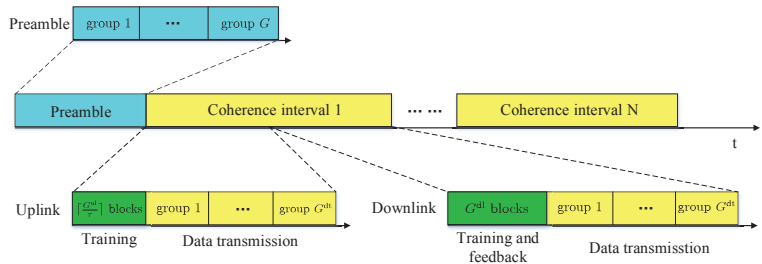}
     \caption{The communication process of the new framework. {\it preamble} is used to collect the information
     about DOAs, then followed by uplink/downlink training and data transmission.}
     \label{fig:trainingphrase}
\end{figure}

We propose a new uplink/downlink transmission framework that utilizes the spatial signatures to realize the orthogonal training and
data transmission among  different users. As shown in  Fig.~\ref{fig:trainingphrase}, the
transmissions between BS and users always start from an uplink preamble to obtain
the spatial signature of each user. Then users are grouped and scheduled for
uplink/downlink training and data transmission based on their spatial signatures such that the orthogonal transmission is achieved.

It is worth mentioning that based on the proposed SBEM, once the spatial signatures of users are obtained in the preamble,
the reduced-dimensional channels can be estimated through traditional linear LS method. However, to make the proposed strategy complete
and address some ideas in detail, we will generally show the whole procedures of proposed channel estimation.

\subsection{Obtain  Spatial Information through Uplink Preamble}

For the ease of illustration, assume $K=G\tau$ where $G\geq 1$ is an integer.
Following most standards, there exists a relative long uplink training period called \emph{preamble} at the very beginning of any transmission, while the functionality of the preamble in the proposed framework  is  mainly to obtain the spatial signatures of different users.
Since we do not assume any prior  spatial information, we will have to divide $K$ users into
$G$ groups, each containing $\tau$ users such that $\tau$ orthogonal training sequences is sufficient for each group. Then, the conventional uplink channel estimation will be applied for each group during the preamble period, and hence the length of the preamble is $GL$.

Taking the first group as an example, the received training signals at BS is given by
\begin{equation}\label{equ:preambletraining}
\begin{aligned}
  \mathbf Y&=\mathbf H\mathbf D^{1/2}\mathbf S^H+\mathbf N
  =\sum_{i=1}^\tau \sqrt{d_i}\mathbf h_{i}\mathbf s_{i}^H+\mathbf N,
\end{aligned}
\end{equation}
where $\mathbf H=[\mathbf h_{1},\cdots,\mathbf h_{\tau}]\in\mathbb{C}^{M\times \tau}$ is the uplink channel
matrix from the first $\tau$ users;
 $\mathbf S\triangleq[\mathbf s_{1},\cdots,\mathbf s_{\tau}]\in\mathbb{C}^{L\times \tau}$
 contains $\tau$ training sequences assigned to this cell; $\mathbf D\triangleq\tu{diag}\{d_1,\ldots,d_{\tau}\}$
 and $d_k\triangleq\frac{P_k^{\tu{ut}}}{L\sigma_p^2}$ is used to satisfy the uplink training energy constraint $P_k^{\tu{ut}}$ of user-$k$;
 $\mathbf N$ is
the independent additive white Gaussian noise matrix with elements distributed as i.i.d
$\mathcal{CN}(0,\sigma_n^2)$.
Then $\mathbf{h}_k$  can be estimated through LS as
\begin{equation}\label{equ:preambletraininghk}
\begin{aligned}
  \hat{\mathbf h}_{k}&=\frac{1}{\sqrt{d_k}L\sigma_p^2}\mathbf Y\mathbf s_{k}=\mathbf h_{k}+\frac{1}{\sqrt{P_k^{\tu{ut}}}/\sigma_n^2}\mathbf n_{k},
\end{aligned}
\end{equation}
where  $\mathbf n_{k}\sim\mathcal{CN}(\mathbf 0,\mathbf I)$ denotes the normalized Gaussian white noise vector.

Repeating the similar operations in \eqref{equ:preambletraininghk} for all $G$ groups
yields the channel estimates for all $K$ users. The next step is   to obtain the optimal rotation $\phi_k$ and extract the spatial signature $\mathcal{B}_k^{\tu{ro}}$ of size $\tau$  that contains the maximum power of $\mathbf F\mathbf{\Phi}(\phi_k)\hat{\mathbf h}_k$, as described in \eqref{equ:phaserotationobjective}.
Then,  the channel vector can be approximated by
\begin{equation}\label{equ:uplinkBEM}
  \mathbf h_k=\mathbf \Phi(\phi_k)^H\mathbf F^H\tilde{\mathbf h}_k^{\tu{ro}}\approx \mathbf \Phi(\phi_k)^H
  \left[\mathbf F^H\right]_{:,\mathcal{B}_k^{\tu{ro}}}\left[\tilde{\mathbf h}_k^{\tu{ro}}\right]_{\mathcal{B}_k^{\tu{ro}},:}.
\end{equation}

\begin{remark}
Contrast to existing works  \cite{yin,Caire}  where channel covariances are directly assumed to be known, we here design a concrete method to acquire the  spatial information  for all users.
\end{remark}

\subsection{Uplink Training with User Grouping}

Channel information obtained from preamble may only last for a short period, say one coherent time, while it
should be tracked or estimated again in the later transmission. Since the channel coherence time is in the level of millisecond while a user and its surrounding obstacles may not
physically change its position in the comparable time, we may treat   DOA and AS of a user as unchanged within several or even tens of the channel
coherence times. Hence, the spatial signatures  $\mathcal{B}_k^{\tu{ro}}$
of each user could be deemed as unchanged, and only the accompanied
$\left[\tilde{\mathbf h}_{k}^{\tu{ro}}\right]_{\mathcal{B}_{k}^{\tu{ro}},:}$ should be re-estimated in the later transmission.
Moreover, the non-overlapping properties of different spatial signatures could also be utilized to release the pressure of insufficient orthogonal training sequences.

Let us then divide users into different groups according
to their spatial signatures.
Specifically, users are allocated to the same group if their spatial signatures do not
overlap and are separated by  a certain guard interval $\Omega$, i.e.,
    \begin{equation}\label{equ:ULgroupingcriterion}
            \mathcal{B}_k^{\tu{ro}}\cap\mathcal{B}_l^{\tu{ro}}=\emptyset,
            \quad \text{and} \quad
        \tu{dist}\left(\mathcal{B}_k^{\tu{ro}},\mathcal{B}_l^{\tu{ro}}\right)\geq \Omega,
    \end{equation}
where $\tu{dist}(\mathcal{B}_1,\mathcal{B}_2)\triangleq\min |b_1-b_2|,\ \forall b_1\in\mathcal{B}_1,\forall b_2\in\mathcal{B}_2$,
 and the value of $\Omega$  depends on the tolerance of users for the interference due to
pilot reusing.

Assume that all users are  divided into $G^{\tu{ut}}$ groups based on \eqref{equ:ULgroupingcriterion} and denote the user index set of the $g$-th group as $\mathcal{U}_g^{\tu{ut}}$. We expect that $G^{\tu{ut}}\ll K$ since $\tau=|\mathcal{B}_k^{\tu{ro}}|$ is much smaller than $M$ and users are randomly distributed in the service region.

As channels of different users in the same group could be discriminated by their spatial signatures, we could assign the same training sequence to all users in one group. Let us first consider the case $G^{\tu{ut}}\leq \tau$ and allocate $\mathbf s_i$ to the $i$-th group. Then, all $K$ users in $G^{\tu{ut}}$ groups send their training sequences simultaneously, and the received signals at BS can be expressed as
    \begin{equation}\label{equ:uplinktraining}
    \begin{aligned}
      \mathbf Y&=\sum_{i=1}^{G^{\tu{ut}}}\sum_{k\in\mathcal{U}_i^{\tu{ut}}}\sqrt{d_i}\mathbf h_k\mathbf s_i^H+\mathbf N.
    \end{aligned}
    \end{equation}
Since $\mathbf s_i$'s are orthogonal to each other, we can extract the signals for group-$g$ as
\begin{align}\label{equ:uplinktrainingyk}
      \mathbf y_{g}&=\frac{1}{L\sigma_{p}^2}\mathbf Y\mathbf s_{g}=
     \frac{1}{L\sigma_{p}^2}\left(\sum_{i=1}^{G^{\tu{ut}}}\sum_{k\in\mathcal{U}_{i}^{\tu{ut}}}\sqrt{d_k}\mathbf h_k\mathbf s_{i}^H+\mathbf N\right)\mathbf s_{g}\nonumber\\
      &=\sum_{l\in\mathcal{U}_{g}^{\tu{ut}}}\sqrt{d_l}\mathbf h_l+\frac{1}{L\sigma_{p}^2}\mathbf N\mathbf s_{g}=\sum_{l\in\mathcal{U}_{g}^{\tu{ut}}}\sqrt{d_l}\mathbf h_l+\frac{1}{L\sigma_p^2/\sigma_n^2}\mathbf{\bar{n}}_{g},
\end{align}
where $\mathbf{\bar{n}}_{g}\sim\mathcal{CN}(\mathbf 0,\mathbf I)$ is the normalized Gaussian white noise vector. Clearly, $\mathbf y_g$ only contains the channel vectors of users in group-$g$ while the interference from  other groups are completely
eliminated by the orthogonal training. The term $\sum_{l\in\mathcal{U}_{g}^{\tu{ut}}}\sqrt{d_l}\mathbf h_l$ in \eqref{equ:uplinktrainingyk} is conventionally called  as \emph{pilot contamination} \cite{marzetta2010noncooperative},  which
is caused by the same training sequence from different users.
Nevertheless,  since different channel vectors in  group-$g$ have different  spatial signatures $\mathcal{B}_k^{\tu{ro}}$, we could  still extract the individual channel information of each user.
For example for  user-$k$ in group-$g$, let us formulate
\begin{align}\label{equ:uplinkyktilde}
     \tilde{\mathbf y}_{g,k}^{\tu{ro}}&=\frac{1}{\sqrt{d_k}}\mathbf F\mathbf \Phi(\phi_k)\mathbf y_{g}=\mathbf F\mathbf \Phi(\phi_k)\mathbf h_k+\sum_{l\in\{\mathcal{U}_{g}^{\tu{ut}}\backslash k\}}
      \sqrt{\frac{d_l}{d_k}}\mathbf F\mathbf \Phi(\phi_k)\mathbf h_l+
      \frac{1}{\sqrt{P_k^{\tu{ut}}/\sigma_n^2}}\mathbf F\mathbf \Phi(\phi_k)\bar{\mathbf n}_{g}\nonumber\\
      &=\tilde{\mathbf h}_k^{\tu{ro}}+\sum_{l\in\{\mathcal{U}_{g}^{\tu{ut}}\backslash k\}}\sqrt{\frac{d_l}{d_k}}\mathbf F\mathbf\Phi(\phi_k){\mathbf h}_l+
      \frac{1}{\sqrt{P_k^{\tu{ut}}/\sigma_n^2}}\mathbf F\mathbf \Phi(\phi_k)\bar{\mathbf n}_{g}.
\end{align}
From SBEM we know the  $\tilde{\mathbf h}_k^{\tu{ro}}$ can be approximated by  $\left[\tilde{\mathbf h}_k^{\tu{ro}}\right]_{\mathcal{B}_k^{\tu{ro}},:}$, while the estimate of  $\left[\tilde{\mathbf h}_k^{\tu{ro}}\right]_{\mathcal{B}_k^{\tu{ro}},:}$  can be extracted from (\ref{equ:uplinkyktilde}) as
\begin{align}\label{equ:partialBk}
    \widehat{\left[\tilde{\mathbf h}_k^{\tu{ro}}\right]}_{\mathcal{B}_k^{\tu{ro}},:}&=\left[\tilde{\mathbf y}_{g,k}^{\tu{ro}}\right]_{\mathcal{B}_k^{\tu{ro}},:}
      =\frac{1}{\sqrt{d_k}}\left[\mathbf I_{{M}}\right]_{\mathcal{B}_k^{\tu{ro}},:}
      \mathbf F\mathbf \Phi(\phi_k)\mathbf y_{g}\notag\\
      &=\left[\tilde{\mathbf h}_k^{\tu{ro}}\right]_{\mathcal{B}_k^{\tu{ro}},:}+
      \sum_{l\in\{\mathcal{U}_{g}^{\tu{ut}}\backslash k\}}\sqrt{\frac{d_l}{d_k}}\left[\mathbf F\mathbf\Phi(\phi_k){\mathbf h}_l\right]_{\mathcal{B}_k^{\tu{ro}},:}
      +\frac{1}{\sqrt{P_k^{\tu{ut}}/\sigma_n^2}}\left[\mathbf F\mathbf \Phi(\phi_k)\bar{\mathbf n}_{g}\right]_{\mathcal{B}_k^{\tu{ro}},:}.
\end{align}
    According to \emph{Property \ref{remark:nonoptimalphase}} and
  bearing  in mind that $\mathcal{B}_l^{\tu{ro}}$ and $\mathcal{B}_k^{\tu{ro}}$ are separated  at least by one guard interval $\Omega$,
  we know the entries of $\left[\mathbf F\mathbf\Phi(\phi_k)\mathbf h_l\right]_{\mathcal{B}_k^{\tu{ro}},:}$
  in \eqref{equ:partialBk} is very small.
Then, the estimate of $\tilde{\mathbf h}_k^{\tu{ro}}$ could be approximated as
    \begin{align}
     \hat{\tilde{\mathbf h}}_k^{\tu{ro}}=\left[\mathbf 0^T\ \widehat{\left[\tilde{\mathbf h}_k^{\tu{ro}}\right]}^H_{\mathcal{B}_k^{\tu{ro}},:}\
      \mathbf 0^T\right]^H,
    \end{align}
    where the two zero vectors $\mathbf 0$ have appropriate sizes.
    Hence, the channel estimate of user-$k$ could be computed as
    \begin{align}\label{equ:uplinktraininghk}
      \hat{\mathbf h}_k&=\mathbf{\Phi}(\phi_k)^H\mathbf F^H\hat{\tilde{\mathbf h}}_k^{\tu{ro}}=\mathbf \Phi(\phi_k)^H\left[\mathbf F^H\right]_{:,\mathcal{B}_k^{\tu{ro}}}\left[\mathbf I_{\textup{M}}\right]_{\mathcal{B}_k^{\tu{ro}},:}
      \mathbf F\mathbf \Phi(\phi_k)\mathbf y_{g}.
    \end{align}
\begin{remark}
The  operation in \eqref{equ:partialBk} to get $\left[\tilde{\mathbf y}_{g}^{\tu{ro}}\right]_{\mathcal{B}_k^{\tu{ro}},:}$ from
    $\mathbf y_g$ could be accelerated by  taking the partial FFT \cite{partialFFT1,partialFFT2,partialFFT3} rather than a full FFT operation, where the number of complex multiplications to get the $\tau$ DFT points is as low as ${\it{O}}(\frac{M}{2}\log_2\tau)$ when $\tau$ is a power of two.
\end{remark}

    The mean square error (MSE) of $\hat{\mathbf h}_k$ in \eqref{equ:uplinktraininghk}
    can be expressed as
    \begin{align}\label{equ:uplinktrainingerror}
      \textup{MSE}_{k}^{\tu{u}}=&\mathbb{E}\left\{\left\|\mathbf h_k-\hat{\mathbf h}_k\right\|^2\right\}=\mathbb{E}\left\{\left\|\mathbf F\mathbf \Phi(\phi_k)\mathbf h_k-\mathbf F\mathbf \Phi(\phi_k)\hat{\mathbf h}_k\right\|^2\right\}\notag\\
    =&\left\|\left[\tilde{\mathbf h}_k^{\tu{ro}}\right]_{\Xi\backslash\mathcal{B}_k^{\tu{ro}},:}\right\|^2+ \left\|\sum_{l\in\mathcal{U}_{g}^{\tu{ut}}\backslash \{k\}}\left[\mathbf F\mathbf\Phi(\phi_k)\mathbf h_l\right]_{\mathcal{B}_k^{\tu{ro}},:}\right\|^2
        +\frac{\tau}{P_k^{\tu{ut}}/\sigma_n^2},
    \end{align}
where $\Xi$ denotes the complete index set $\{0,\ldots,M-1\}$.

It is easy to infer from \eqref{equ:uplinktrainingerror} that the channel estimation
error is composed of three parts.  The first part is obviously the truncation error from
SBEM that only keeps $\tau$ DFT points in $\tilde{\mathbf h}^{\tu{ro}}_k$. The channel leakage from other users in the same group to the current $\mathcal{B}_k^{\tu{ro}}$ brings the second
error term, which formulates \emph{remaining} pilot contamination. The noise term accounts for
the last error part, which is proportional to $\tau$ \cite{LSmse}.
The truncation error will be reduced if  AS of user-$k$ becomes smaller or if the value of $\tau$ increases. However, AS is the inherent property of the environment and $\tau$  is generally  regulated by
the standards, both being most likely uncontrollable.  Nevertheless, the inter-user interferences can be  reduced when $\Omega$ is set to be larger. Hence, if $G^{\tu{ut}}$ is smaller than $\tau$,  then we should reformulate exactly $G^{\tu{ut}}=\tau$ groups with evenly distributed users in each group such that the second error term can be reduced.

On the other side, if $G^{\tu{ut}}>\tau$, then $\tau$ orthogonal training sequences cannot be exclusively assigned to different user groups.
In this case, we have to train  $\tau$ groups after another $\tau$  groups  until all $G^{\tu{ut}}$ groups complete the channel estimation period, while the training
approach for each $\tau$ groups is the same as discussed above.


\subsection{Angle Reciprocity and Downlink Channel Representation}\label{sec:downlinktraining}
 Denote the downlink channel from BS to  user-$k$ as $\mathbf g_k^H\in\mathbb{C}^{1\times M}$.
 Similar to  \eqref{equ:channelmodel},
 $\mathbf g_k\in\mathbb{C}^{M\times 1}$ can be modeled as
 \begin{align}
   \mathbf g_k=\frac{1}{\sqrt{P}}\sum_{p=1}^P\beta_{kp}\mathbf a(\vartheta_{kp}),
 \end{align}
where $\mathbf a(\vartheta_{kp})$ is the steering vector defined in \eqref{equ:steeringvector} but with different downlink carrier wavelength
 $\lambda_2$; $\vartheta_{kp}$ is the DOD of the $p$-th ray that could arrive at user-$k$, and  $\beta_{kp}$ is the corresponding
 complex gain. All other parameters have the same definitions as in \eqref{equ:channelmodel}.

Similar to the uplink, we expect that the downlink channel $\mathbf g_k$ can also be approximated  by SBEM with appropriate
spatial signatures $\overline{\mathcal{B}_k^{\textup{ro}}}$ and phase shift parameter $\bar{\phi}_k$, i.e.,
\begin{align}\label{equ:downlinkbeamchannel}
  \mathbf g_{k}&=\mathbf \Phi(\bar{\phi}_k)^H\mathbf F^H\tilde{\mathbf g}_{k}^{\textup{ro}}
\approx\mathbf \Phi(\bar{\phi}_k)^H\left[\mathbf F^H\right]_{:,\overline{\mathcal{B}_k^{\textup{ro}}}}
\left[\tilde{\mathbf g}_{k}^{\textup{ro}}\right]_{\overline{\mathcal{B}_k^{\textup{ro}}},:}
=\sum_{q\in\overline{\mathcal{B}_k^{\textup{ro}}}}\tilde{g}_{k,q}^{\textup{ro}}\mathbf \Phi(\bar{\phi}_k)^H\mathbf f_q,
\end{align}
where $\tilde{g}_{k,q}^{\textup{ro}}\triangleq[\tilde{\mathbf g}_k^{\textup{ro}}]_q$ denotes the $q$-th element
of $\tilde{\mathbf g}_k^{\textup{ro}}=\mathbf F\mathbf \Phi(\bar{\phi}_k)\mathbf g_k$.
Following \eqref{equ:downlinkbeamchannel}, once $\overline{\mathcal{B}_k^{\textup{ro}}}$ and $\bar{\phi}_k$ are
determined, the estimation of downlink channels $\mathbf{g}_k$ is simplified to estimate the
unknown SBEM coefficients $\tilde{g}_{k,q}^{\textup{ro}}$.

Since the propagation path of electromagnetic wave is reciprocal, we know that only the signal
wave that physically reverses the uplink path can reach the user during the downlink period.
Hence, downlink signals that could effectively arrive at the user should have the same DOD spread as the uplink DOA spread,
namely, the angle range of $\vartheta_{kp}$'s is the same as the angle range of $\theta_{kp}$'s. We call
this property of the wireless channel as \emph{angle reciprocity}.\footnote{If
 the frequency of the downlink channel is  not too far from that of  the uplink channel, e.g. less than several GHz,  then the reciprocity  between DOD and DOA accurately holds. The reason is that for the typical transmission environment, the relative permittivity and the conductivity of the obstacles do not change
in the scale of  several dozens of GHz \cite{EMproperty}. Hence, the reflection and the deflection properties of signals
with  less than several GHz frequency differences is almost identical in  such propagation environment \cite{winner,metis}.} For TDD system, the channel gain is also reciprocal such that the overall
channel is reciprocal.

\begin{remark}
Angle reciprocal may not be that useful in conventional MIMO system but is specifically valuable  for massive MIMO system where the channel estimation
can be decomposed into gain estimation and angle estimation.
\end{remark}

 Based on the angle reciprocity,  $\overline{\mathcal{B}_k^{\textup{ro}}}$ can be  determined by $\mathcal{B}_k^{\textup{ro}}$.
Specifically, according to \emph{Property \ref{property:2}}, we will have
\begin{align}
\sin\theta_{kp} = \frac{q\lambda_1}{Md}=\frac{q'\lambda_2}{Md},\ \textup{with}\ q\in\mathcal{B}_k^{\textup{ro}},\
q'\in\overline{\mathcal{B}_k^{\textup{ro}}},
\end{align}
where $\lambda_1$ and $\lambda_2$ denote the uplink/downlink carrier wavelengths, respectively.
\textcolor[rgb]{0.00,0.00,1.00}{Then the integer set $\overline{\mathcal{B}_k^{\textup{ro}}}$ can
be expressed as $\overline{\mathcal{B}_k^{\textup{ro}}}=\{q'_{\min},q'_{\min}+1,\ldots,q'_{\max}\}$ with
\begin{align}\label{equ:downlinkBk}
  q'_{\min}=\left\lfloor\frac{\lambda_1}{\lambda_2}q_{\min}\right\rfloor,\ \ q'_{\max}=\left\lceil\frac{\lambda_1}{\lambda_2}q_{\max}\right\rceil,
\end{align}}
where $q_{\min}\leq q\leq q_{\max}, \forall q\in\mathcal{B}_k^{\textup{ro}}$.
Similarly, $\bar{\phi}_k$ can be determined as $\bar{\phi}_k = \frac{\lambda_1}{\lambda_2}\phi_k$.

\begin{remark}
A key advantage of the proposed antenna array theory based approach over the covariance matrix based method \cite{yin}, \cite{Caire} is that the angle of the downlink FDD channel can be predicted
from the uplink channel and can be used to simplify the downlink channel estimation.
\end{remark}

\subsection{Downlink Training with User Grouping}\label{sec:downlinkgrouping}
The key difficulty to apply the conventional downlink channel estimation algorithms for massive MIMO system lies in the requirement that
the length of the training has to be no less than the number of antennas as well as the high computational complexity
when computing massive parameters. Moreover, the feedback of huge
channel state information from user to BS also costs severe overhead.

With  SBEM in \eqref{equ:downlinkbeamchannel}, downlink channel estimation for each user only needs to estimate $\tau$  parameters.
To reuse the overall $\tau$ orthogonal training sequences, let us divide $K$ users into different groups. We first gather users with identical spatial signature  $\overline{\mathcal{B}_k^{\textup{ro}}}$
into the same cluster.
Then, we start to assign clusters into different groups such that the spatial signatures of the clusters in the same group do not overlap and
are separated by a certain guard interval, as did in  \eqref{equ:ULgroupingcriterion}.
For the ease of exposition, assume that all $K$ users are divided into $G^{\tu{dt}}$ groups and
denote the user index set of the $g$-th group as $\mathcal{U}_g^{\tu{dt}}$.

Let us then take the training of the group-$g$ for example to describe the downlink channel estimation algorithm.
Based on SBEM, the effective transmission between $\tau$ DFT points in   $\overline{\mathcal{B}_k^{\textup{ro}}}$ and each user formulates a virtual $
\tau \times  1$ MISO downlink system. Hence, to estimate the $\tau$ coefficients $[\tilde{\mathbf g}_k^{\tu{ro}}]_{\overline{\mathcal{B}_k^{\textup{ro}}}}$ for each user, we need to transmit
$\tau$ orthogonal training sequences from the corresponding $\tau$ beams $\mathbf f_q,\ q\in\overline{\mathcal{B}_k^{\textup{ro}}}$.
We then select the orthogonal training matrix for user-$k$
as $\mathbf S_k=\varpi_k\mathbf S^H\in\mathbb{C}^{\tau\times L}$, where the
scalar variable $\varpi_k$ is used to satisfy the transmit power constraint $\tu{tr}\{\mathbf S_k\mathbf S_k^H\}\leq P_k^{\tu{dt}}$,
and thus $\varpi_k=\sqrt{\frac{P_k^{\tu{dt}}}{\tau L\sigma_p^2}}$.
Note that, the same orthogonal training matrix
$\mathbf S^H\in\mathbb{C}^{\tau\times L}$ is reused by different users over their own spatial signatures in the same group.

Then the received signals at  user-$k$ of group-$g$ can be expressed as
\begin{align}\label{equ:dltraingforuserk}
    \mathbf y_{k}^H&=\mathbf g_{k}^H
    \left(\sum_{l\in\mathcal{U}_g^{\tu{dt}}}\mathbf \Phi(\bar{\phi}_l)^H\left[\mathbf F^H\right]_{:,\overline{\mathcal{B}_l^{\tu{ro}}}}\mathbf S_l\right)+\mathbf n_k^H=\sum_{l\in\mathcal{U}_g^{\tu{dt}}}\left(\left[\mathbf F\right]_{\overline{\mathcal{B}_l^{\tu{ro}}},:}\mathbf \Phi(\bar{\phi}_l)\mathbf g_{k}\right)^H \mathbf S_l+\mathbf n_k^H\notag\\
    &= \left[\tilde{\mathbf g}_{k}^{\tu{ro}}\right]^H_{\overline{\mathcal{B}_k^{\tu{ro}}},:}\mathbf S_k+
    \sum_{l\in\{\mathcal{U}_g^{\tu{dt}}\backslash k\}}\left[\mathbf F\mathbf \Phi(\bar{\phi}_l)\mathbf g_{k}\right]_{
    \overline{\mathcal{B}_l^{\tu{ro}}},:}^H \mathbf S_l
    +\mathbf n_k^H,
\end{align}
where  $\mathbf n_k^H\in\mathbb{C}^{1\times L}$
is the noise vector with its elements distributed as  $\mathcal{CN}(0,\sigma_{n}^2)$.
The downlink channel of user-$k$ can be estimated through LS method as
\begin{align}\label{equ:downlinkLS}
  \widehat{\left[\tilde{\mathbf g}_{k}^{\tu{ro}}\right]}^H_{\overline{\mathcal{B}_k^{\textup{ro}}},:}=&
\mathbf y_{k}^H\mathbf S_k^\dag
= \left[\tilde{\mathbf g}_{k}^{\tu{ro}}\right]^H_{\overline{\mathcal{B}_k^{\textup{ro}}},:}+
    \sum_{l\in\{\mathcal{U}_g^{\tu{dt}}\backslash k\}}\frac{\varpi_l}{\varpi_k}\left[\mathbf F\mathbf \Phi(\bar{\phi}_l)\mathbf g_{k}\right]_{\overline{\mathcal{B}_l^{\tu{ro}}},:}^H +\mathbf n_k^H\mathbf S_k^\dag,
\end{align}
where the second term is the
interference coming from the DFT points  $\mathbf F\mathbf \Phi(\bar{\phi}_l)\mathbf g_{k}$ in set $\overline{\mathcal{B}_l^{\tu{ro}}}, l\in \mathcal{U}_g^{\textup{dt}}, l\neq k$, and is a kind of
self-interference.\footnote{There is no pilot contamination in the downlink training,
but the reusing of ${\mathbf S^H}$ along different spatial signatures  will introduce the
self-interference.}
Recalling \emph{Property \ref{remark:nonoptimalphase}}, since
$[\mathbf F\mathbf \Phi(\bar{\phi}_l)\mathbf g_{k}]_{\overline{\mathcal{B}_k^{\tu{ro}}},:}$ contains most
channel energy and since $\overline{\mathcal{B}_l^{\textup{ro}}}$ is
separated at least one guard interval from $\overline{\mathcal{B}_k^{\textup{ro}}}$,
$\left[\mathbf F\mathbf \Phi(\bar{\phi}_l)\mathbf g_{k}\right]_{\overline{\mathcal{B}_l^{\tu{ro}}},:}$
will be very small compared to $\left[\tilde{\mathbf g}_{k}^{\tu{ro}}\right]^H_{\overline{\mathcal{B}_k^{\textup{ro}}},:}$.

Then the estimate of $\tilde{\mathbf g}_k^{\tu{ro}}$ can be expressed as
\begin{align}
  \hat{\tilde{\mathbf g}}_k^{\tu{ro}}=\left[\mathbf 0^T\ \widehat{\left[\tilde{\mathbf g}_k^{\tu{ro}}\right]}^H_{\overline{\mathcal{B}_k^{\textup{ro}}},:}\ \mathbf 0^T\right]^H,
\end{align}
where the two zero vectors $\mathbf 0$ have appropriate sizes.
Following \eqref{equ:downlinkbeamchannel}, the channel estimate for user-$k$ in group-$g$ is obtained as
\begin{align}\label{equ:downlinkchannelestimation}
  \hat{\mathbf g}_k=\mathbf \Phi(\bar{\phi}_k)^H\mathbf F^H\hat{\tilde{\mathbf g}}_k^{\tu{ro}}=
  \mathbf \Phi(\bar{\phi}_k)^H\left[\mathbf F^H\right]_{:,\overline{\mathcal{B}_k^{\textup{ro}}}}\left(\mathbf S_k^H\right)^\dag\mathbf y_k.
\end{align}

After channel estimation,   each user feeds back $\tau$ components $[\hat{\tilde{\mathbf g}}_{k}^{\tu{ro}}]_{\overline{\mathcal{B}_k^{\textup{ro}}},:}$
to BS such that  BS can perform the optimal user scheduling and power allocation for the subsequent downlink data transmission.  Compared to the feedback of large amount of measurements in \cite{KLT}, the overhead of the newly proposed framework
is significantly reduced.

Similar to the analysis of uplink, the downlink MSE of the LS estimator for user-$k$
can be expressed as
\begin{align}\label{equ:dlmse}
    \textup{MSE}_{k}^{\tu{d}}=&\mathbb{E}\left\{\left\|\mathbf g_{k}-\hat{\mathbf g}_{k}\right\|^2\right\}
            =\mathbb{E}\left\{\left\|\tilde{\mathbf g}_{k}^{\tu{ro}}-\tilde{\hat{\mathbf g}}_{k}^{\tu{ro}}\right\|^2\right\}=\left\|\tilde{\mathbf g}_{k}^{\tu{ro}}-\left[\mathbf  0^H\ \left[\hat{\tilde{\mathbf  g}}_{k}^{\tu{ro}}\right]^H_{\overline{\mathcal{B}_k^{\textup{ro}}},:}\ \mathbf  0^H\right]^H\right\|^2\notag\\
            =&\left\|\left[\tilde{\mathbf g}_{k}^{\tu{ro}}\right]_{\Xi\backslash\overline{\mathcal{B}_k^{\textup{ro}}},:}\right\|^2
            +\left\|\sum_{l\in\{\mathcal{U}_g^{\tu{dt}}\backslash k\}}\frac{\varpi_l}{\varpi_k}\left[\mathbf F\mathbf \Phi(\bar{\phi}_l)\mathbf g_{k}\right]_{\overline{\mathcal{B}_l^{\textup{ro}}},:}^H \right\|^2
            +\frac{\tau^2}{P_k^{\tu{dt}}/\sigma_n^2},
\end{align}
which is  also comprised of three parts. The first part is, again,  the truncation error from SBEM
with only $\tau$ spatial signatures. The channel leakage of SBEM from user-$k$ itself
results in the second term, which is different from  uplink case.
The last error term comes from the noise and  is proportional to
$\tau^2$, which is consistent to the result in the conventional $\tau\times 1$ MISO downlink system \cite{LSmse}.
The self-interference due to pilot reuse will be reduced when the number of  users in the same group decreases and the guard interval becomes larger.

\begin{remark}
From (\ref{equ:downlinkLS}), it is seen that user-$k$ does not
need the knowledge of spatial signature set $\overline{\mathcal{B}_k^{\textup{ro}}}$ and the shift parameter $\bar{\phi}_k$ to perform the estimation of $\tau$ channel  parameters ${\tilde{\mathbf g}}_k^{\tu{ro}}$. This removes the necessity of feedback from BS to the user and is thus a key advantage that make the proposed downlink channel estimation strategy suitable when the mobile users gradually change their positions.
\end{remark}

\section{Data Transmission with User Scheduling}\label{sec:datatransmission}
After obtaining the spatial signatures and  channel gains of all users, we may schedule users into different groups to
enhance the spectral efficiency during the subsequent data transmission period. To achieve this, users in the same group should  have non-overlapping
spatial signatures  such that the inter-user interference could be reduced.  Meanwhile, we try to maximize the achievable rate for each group
under given power constraint.
In the following discussion, we will focus on the downlink case due to page limitation, while the uplink case can
be analyzed in a similar manner.

Assume that users are scheduled into  $G^{\tu{dd}}$ groups for data transmission and
the user index set of the $g$-th group is denoted by
$\mathcal{U}_g^{\tu{dd}}$, $ g=1,2,\ldots,G^{\tu{dd}}$.
For user-$k$ in the $g$-th group, its downlink
received signal can be expressed as
\begin{align}\label{equ:dlreceviedsignal}
 y_{k}^{\tu{dl}}=&\sqrt{\rho_{k}}\mathbf g_{k}^H\mathbf w_{k}x_{k}+\sum_{l\in\{\mathcal{K}_g\backslash k\}}\sqrt{\rho_{l}}\mathbf g_{k}^H\mathbf w_{l}x_{l}+n_{k}\notag\\
  =&\sqrt{\rho_{k}}(\hat{\mathbf g}_{k}^H+\Delta\mathbf g_{k}^H)\mathbf w_{k}x_{k}+\sum_{l\in\{\mathcal{K}_g\backslash k\}}\sqrt{\rho_{l}}
  (\hat{\mathbf g}_{k}^H+\Delta\mathbf g_{k}^H)\mathbf w_{l}x_{l}+n_{k},
\end{align}
 where channel estimation error $\Delta\mathbf g_{k}$ is defined as $\Delta\mathbf g_{k}\triangleq\mathbf g_{k}-\hat{\mathbf g}_{k}$;
$x_{k}$ and $\mathbf w_{k}$ denote the data symbol and the corresponding downlink beamforming vector,
respectively; $\rho_{k}>0$ denotes the transmit power, and $n_{k}\sim\mathcal{CN}(0,1)$ is the received noise.

For downlink massive MIMO systems,  the linear matched filter (MF) beamforming
was already shown to be a good candidate \cite{downlinkuserselection},  namely,
\begin{align}\label{equ:wk}
\mathbf w_{k}=\frac{\hat{\mathbf{g}}_{k}}{\|\hat{\mathbf {g}}_{k}\|^2}=
\frac{\sum_{q\in\mathcal{B}_k^{\tu{ro}}}\hat{\tilde{g}}_{k,q}^{\tu{ro}}\mathbf \Phi(\phi_k)^H\mathbf f_q}{\sum_{q\in\mathcal{B}_k^{\tu{ro}}}|\hat{\tilde{g}}_{k,q}^{\tu{ro}}|^2}.
\end{align}
Interestingly, the structure of $\mathbf{w}_k$ indicates the overall beamforming is composed of $\tau$ orthogonal sub-beams that correspond to $\tau$
spatial signatures of user-$k$, and the beam gains are exactly selected as their estimated channel gains $\hat{\tilde{g}}_{k,q}^{\tu{ro}}$'s.
Since users in the same group have non-overlapped spatial signatures, i.e., $\mathcal{B}_{k}^{\tu{ro}}\cap \mathcal{B}_l^{\tu{ro}}=\emptyset$,
it directly leads to $\mathbf w_{k}^H\mathbf w_{l}=\hat{\mathbf{g}}_k^H\hat{\mathbf {g}}_l=0$. Note that, the beam orthogonality in our proposed strategy always holds for any
finite value of $M$, as contrast to many other works that require $M\rightarrow \infty$ \cite{yin,CCE,meixia}.

\begin{algorithm}[t]
\caption{: User Scheduling Algorithm for Data Transmission}
\label{alg::userscheduling}
\begin{itemize}
\item \textbf{Step 1:} Calculate the Euclidean norm of the estimated channel vectors,
       i.e., $\left\|\hat{\mathbf g}_l\right\|=\left\|\left[\hat{\tilde{\mathbf g}}_{l}^{\tu{ro}}\right]_{\mathcal{B}_l^{\tu{ro}},:}\right\|$, for all users.
\item \textbf{Step 2:} Initialize $g=1$, $\mathcal{P}=0$, $\mathcal{U}_g^{\tu{dd}}=\emptyset$,
        \ $R(\mathcal{U}_g^{\tu{dd}}|\mathcal{P})=0$, and
       the remaining user set $\mathcal{U}_r=\{1,\ldots,K\}$.
\item \textbf{Step 3:} For  the $g$-th group, select the user in $\mathcal{U}_r$ with the maximal norm of channel,
       $l'=\mathop{\argmax}_{l\in \mathcal{U}_r}\left\|\hat{\mathbf g}_{l}\right\|$.
       Set  $\mathcal{P}=\rho$, $\mathcal{U}_g^{\tu{dd}}=\mathcal{U}_g^{\tu{dd}}\cup\{l'\}$
       and $\mathcal{U}_r=\mathcal{U}_r\backslash\{l'\}$. Calculate
       $R(\mathcal{U}_g^{\tu{dd}}|\mathcal{P})$ according to \eqref{equ:BFoptimization}.
\item \textbf{Step 4:} Select all those users in $\mathcal{U}_r$  whose spatial signatures are non-overlapping
       with  users in $\mathcal{U}_g^{\tu{dd}}$, and denote them by $\mathcal{U}'_g$, i.e.,
       \begin{equation*}
           \mathcal{U}'_g=\{m\in\mathcal{U}_r\ |\ \mathcal{B}^{\tu{ro}}_m\cap \mathcal{B}^{\tu{ro}}_l=\emptyset\
           \text{and}\ \tu{dist}(\mathcal{B}_m^{\tu{ro}},\mathcal{B}_l^{\tu{ro}})\geq \Omega,\ \forall\ l\in\mathcal{U}_g^{\tu{dd}}\}.
       \end{equation*}
\item \textbf{Step 5:}  If $\mathcal{U}'_g\neq\emptyset$, set $\mathcal{P}'=\mathcal{P}+\rho$,
        and find a user $m'$ in $\mathcal{U}'_g$, such that
       \begin{equation*}
         m'=\mathop{\argmax}_{m\in\mathcal{U}'_g}\  R\left(\mathcal{U}_g^{\tu{dd}}\cup\{m\}|\mathcal{P}'\right).
       \end{equation*}
       If  $R(\mathcal{U}_g^{\tu{dd}}\cup\{m\}|\mathcal{P}')\geq R(\mathcal{U}_g^{\tu{dd}}|\mathcal{P})$, set $\mathcal{U}_g^{\tu{dd}}=\mathcal{U}_g^{\tu{dd}}\cup\{m'\}$,
       $\mathcal{P}=\mathcal{P}'$,\ $\mathcal{U}_r=\mathcal{U}_r\backslash\{m'\}$
       and go to \textbf{Step 4}; Else, go to \textbf{Step 6.}
\item \textbf{Step 6:} Store $\mathcal{U}_g^{\tu{dd}}$ and $R(\mathcal{U}_g^{\tu{dd}}|\mathcal{P})$. If $\mathcal{U}_r\neq\emptyset$, let $g=g+1$,
       go to \textbf{Step 3}; Else, go to \textbf{Step 7.}
\item \textbf{Step 7:} When the algorithm is stopped, the minimal number of user group $G^{\tu{dd}}$ is set as the current $g$,
and the optimal user scheduling result is accordingly given by $\mathcal{U}_1,\ldots,\mathcal{U}_{G^{\tu{dd}}}$.
\end{itemize}
\end{algorithm}

Then \eqref{equ:dlreceviedsignal} can be simplified as
\begin{align}\label{equ:receivedata}
  y_{k}^{\tu{dl}}
=&\sqrt{\rho_{k}}x_{k}+
\Delta\mathbf g_{k}^H\sum_{l\in\mathcal{K}_g}\sqrt{\rho_{l}}
\frac{\hat{\mathbf g}_{l}}{\|\hat{\mathbf g}_{l}\|^2}x_{l}+n_{k}.
\end{align}
Since we do not assume channel statistics, we can not apply the conventional way
 \cite{ULDLantenna} to characterize the lower bound
of achievable throughput. Nevertheless, in practical case the estimated channel $\hat{\mathbf{g}}_k$ will be used as if it is the true value of $\mathbf{g}_k$.
Hence, we will treat the second term in \eqref{equ:receivedata} as gauss noise with
the same covariance, which is a worst case and then solve the following optimization problem to
approximately maximize the throughput of each group:
\begin{equation}\label{equ:BFoptimization}
\begin{aligned}
\max_{\{\rho_{k}\}}\ \ & R(\mathcal{U}_g^{\tu{dd}}|\mathcal{P})\triangleq\sum_{k\in\mathcal{U}_g^{\tu{dd}}}\log_2(1+\rho_k)\\
\tu{s.t.}\ \ \ &\sum_{k\in\mathcal{U}_g^{\tu{dd}}}\frac{\rho_{k}}{\left\|\left[\hat{\tilde{\mathbf g}}_k^{\tu{ro}}\right]_{\mathcal{B}_k^{\tu{ro}},:}\right\|^2}\leq \mathcal{P},
\end{aligned}
\end{equation}
where $\mathcal{P}$ is the total power constraint for this group. Solutions to \eqref{equ:BFoptimization} can be obtained by the standard  water-filling algorithm \cite{water-filling}.

We then provide a greedy user scheduling algorithm, where the user with the strongest channel gain will be first scheduled and then the other
users with non-overlapping spatial signatures can join the same group only if the achievable sum-rate of the whole group
increases afterwards. The detailed steps can be found in Algorithm \ref{alg::userscheduling}, where the power constraint $\mathcal{P}$ for each group is adjusted dynamically and
is proportional to the final number of users in each group. 

\section{Simulations}\label{sec:simulation}
In this section, we demonstrate the effectiveness of the proposed strategy through  numerical examples.
We select $M=128$, $d=\lambda/2$, and consider $K=32$ active users   that are gathered into
$4$ spatially distributed clusters  around  the mean directions of
$\{-48.59^\circ, -14.48^\circ, 14.48^\circ, 48.59^\circ\}$ respectively.
The channel vectors of different users are formulated  according to \eqref{equ:channelmodel},
where $P=100$, and $\alpha_{kp}$ is independently taken from
$\mathcal{CN}(0,1)$ for all rays and all users;  $\theta_{kp}$ is uniformly distributed inside $[\theta_k-\Delta\theta_k,\theta_k+\Delta\theta_k]$, where AS is supposed be $\Delta\theta_k=2^\circ$ for all users. 
The system coherence interval is set as $T=128$ and
the default value of $\tau$ is assumed to be $\tau=16$, which is only $1/8$ of the antenna number,
and the  guard interval for user grouping is set as $\Omega=\tau/4$. The length of pilot $L$ should satisfy $16\leq L\leq 128$ and is then taken as $L=16,\ 32,\ 64$, respectively.
The signal-to-noise ratio (SNR) is defined as $\rho=\sigma_p^2/\sigma_n^2$.
The performance metric of the channel estimation is taken as the normalized
MSE, i.e.,
\begin{equation*}
  \tu{MSE} \triangleq \frac{\sum_{k=1}^K\left\|\mathbf h_k-\hat{\mathbf h}_k\right\|^2}{\sum_{k=1}^K\left\|\mathbf h_k\right\|^2}.
\end{equation*}
In all examples, the spatial information of users are estimated from the preamble. When $\tau=16$, the 32 users are scheduled into
$2$ groups while when $\tau=8$,  the 32 users are scheduled into
$4$ groups, such that the orthogonal training can be applied to obtain the spatial signatures.

\begin{figure}[t]
\centering
\includegraphics[width=100mm]{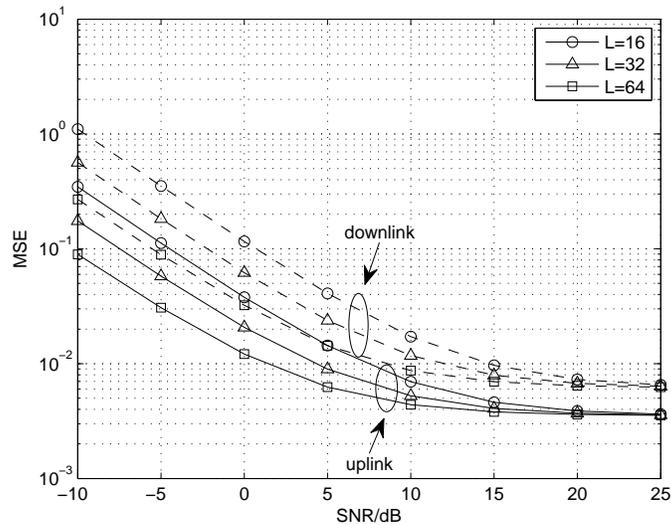}
\caption{Comparison of uplink/downlink MSE performances of the  proposed SBEM method with $\tau=16$ and $L=16,32,64$, respectively.
\label{fig:ULMSE}}
\end{figure}

Fig. \ref{fig:ULMSE} illustrates the MSE performances of uplink/downlink training in \eqref{equ:uplinktraininghk}
and \eqref{equ:downlinkchannelestimation}, respectively, as a function of SNR with different training sequence length $L$.
The total power for both uplink and downlink training is constrained to
$P_k^{\tu{ut}}=P_k^{\tu{dt}}=L\rho$ for all users at a given SNR $\rho$.
For the uplink training, $K=32$ users are divided into $G^{\tu{ut}}=16$ groups.
All these $16$ groups can
be scheduled in the same training length $L$ with $\tau=16$ available orthogonal training sequences. While
for the downlink training, $K=32$ users are gathered into $4$ clusters and are  assigned into $G^{\tu{dt}}=1$ group, i.e., they can be scheduled simultaneously in the same training length $L$ too.
It is seen from  Fig. \ref{fig:ULMSE}  that when  $L$ increases,
the MSE performances of uplink/downlink can be improved, since the total training power is proportional to $L$.
It is also seen that as the SNR increases, there is the same error floor for all values of $L$ during  both uplink/downlink training. This phenomenon is not unexpected due to the truncation error of SBEM from the real channel and  can also be observed in temporal BEM \cite{temporalBEM1}. Meanwhile, since the truncation error is only related with the effective expansion number $\tau$, the error floors will keep
the same for different $L$.
Moreover, it can be seen that the uplink MSE performances are generally better than that of downlink for any SNR and $L$, and
each of uplink MSE curves is almost parallel to the corresponding downlink one with a fixed gap between them. This can be inferred by
comparing the noise terms of \eqref{equ:uplinktrainingerror} and \eqref{equ:dlmse}, where the noise power
included in the uplink training is only proportional to $\tau$ while it is proportional to $\tau^2$ for the downlink training.

\begin{figure}[t]
\centering
\includegraphics[width=100mm]{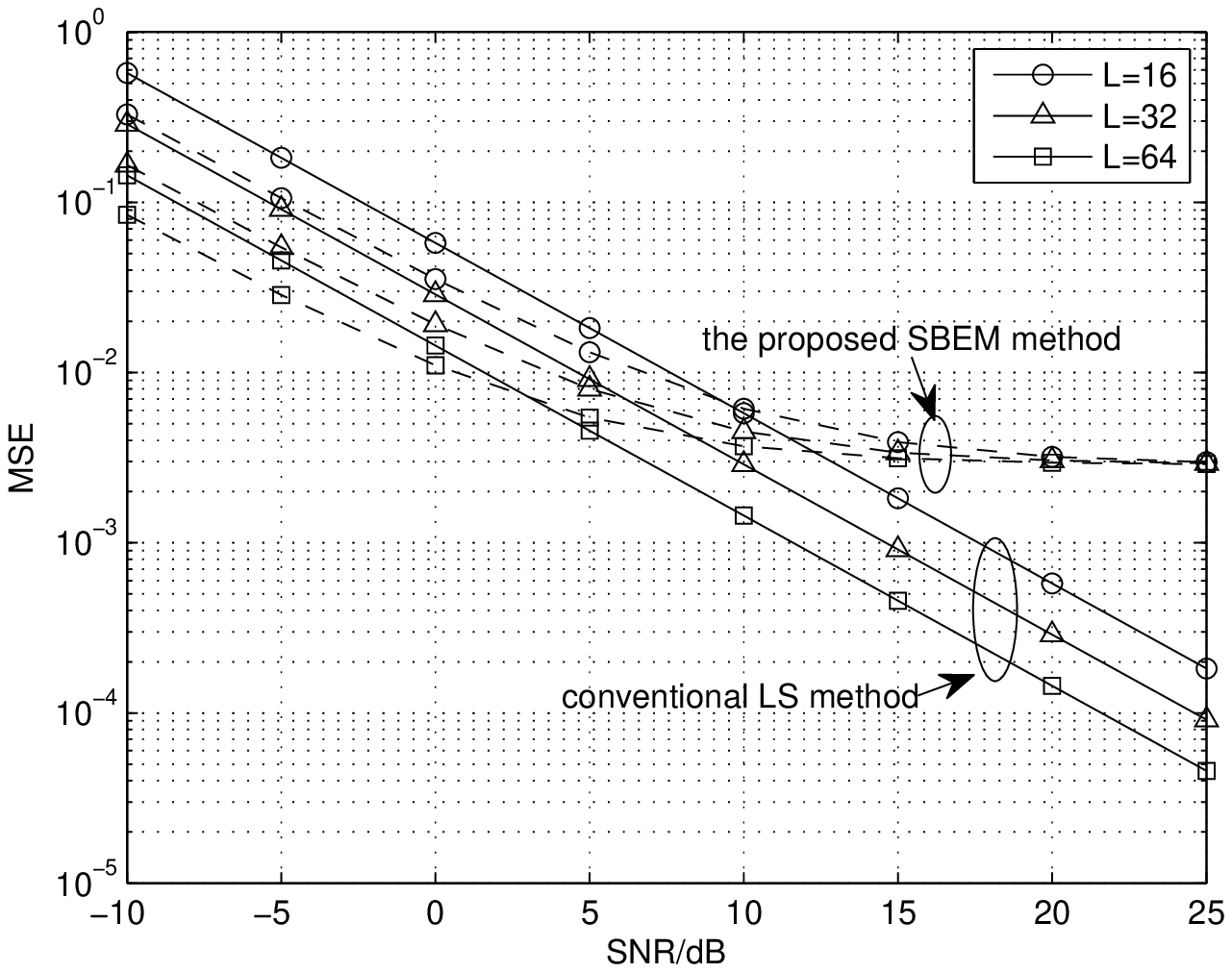}
\caption{The uplink MSE performance comparison of the proposed SBEM method and the conventional LS method, with $\tau=16$ and $L=16,32,64$, respectively.
\label{fig:ULMSEcompare}}
\end{figure}

\begin{figure}[t]
\centering
\includegraphics[width=100mm]{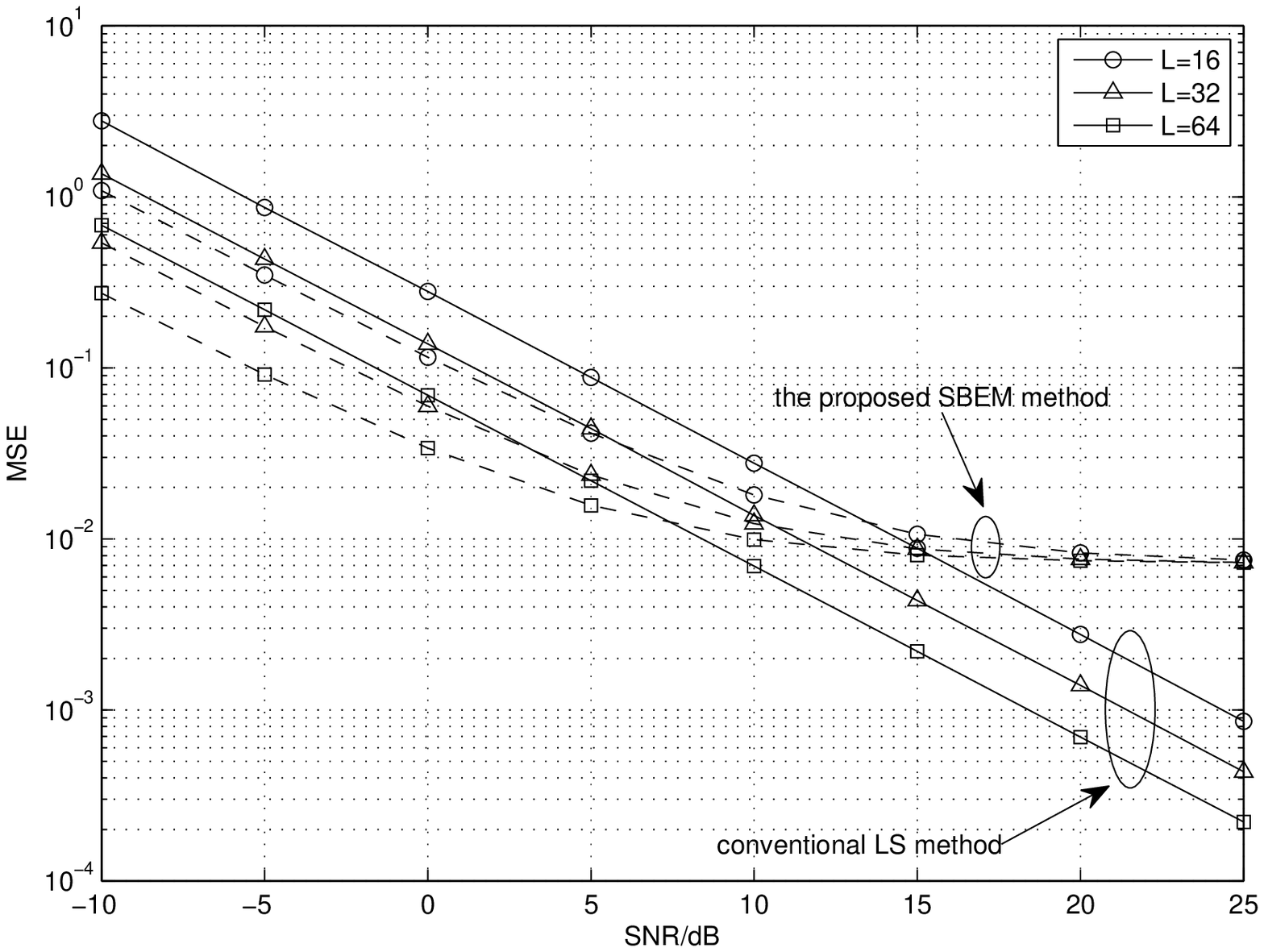}
\caption{The downlink MSE performance comparison of the proposed SBEM method and the conventional LS method, with $\tau=16$ and $L=16,32,64$, respectively.
\label{fig:DLMSEcompare}}
\end{figure}
 Fig. \ref{fig:ULMSEcompare} and Fig. \ref{fig:DLMSEcompare}
compare the proposed channel estimation with the convention LS method for both uplink and downlink cases.
To apply the conventional method, $K=32$ orthogonal training sequences
is used for uplink case while $128\times 128$ orthogonal training matrix is used for downlink case. To provide a fair
comparison, for any given $\rho$ and $L$
the uplink training power is kept the same
$P_k^{\tu{ut}}=L\rho$  for both methods, while the total downlink training power for both methods
are constrained as $\sum_{k=1}^KP_k^{\tu{dt}}=KL\rho$.
Note that, to compared with LS, we set $T=128$, and under such conditions,
LS will consume all $T=128$ symbols for training without any remaining time for data transmission, while
the uplink and downlink training overheads of the proposed scheme are only
$\lceil\frac{G^{\textup{ut}}}{\tau}\rceil L\ll T$ and $G^{\textup{dt}}L\ll T$ respectively.
It is seen that when the sufficient number of orthogonal training is available and
when the computational complexity is acceptable, then the conventional LS method
does not have the error floor for both uplink and downlink cases.
Nevertheless, it is interesting to observe that the channel estimation from SBEM outperforms the conventional method when SNR is relatively low. The reasons can be
found from \eqref{equ:uplinktrainingerror} and \eqref{equ:dlmse} where the proposed method only involves $\tau$ components of the noise vector
while the conventional LS method includes the whole noise power.

\begin{figure}[t]
\centering
\includegraphics[width=100mm]{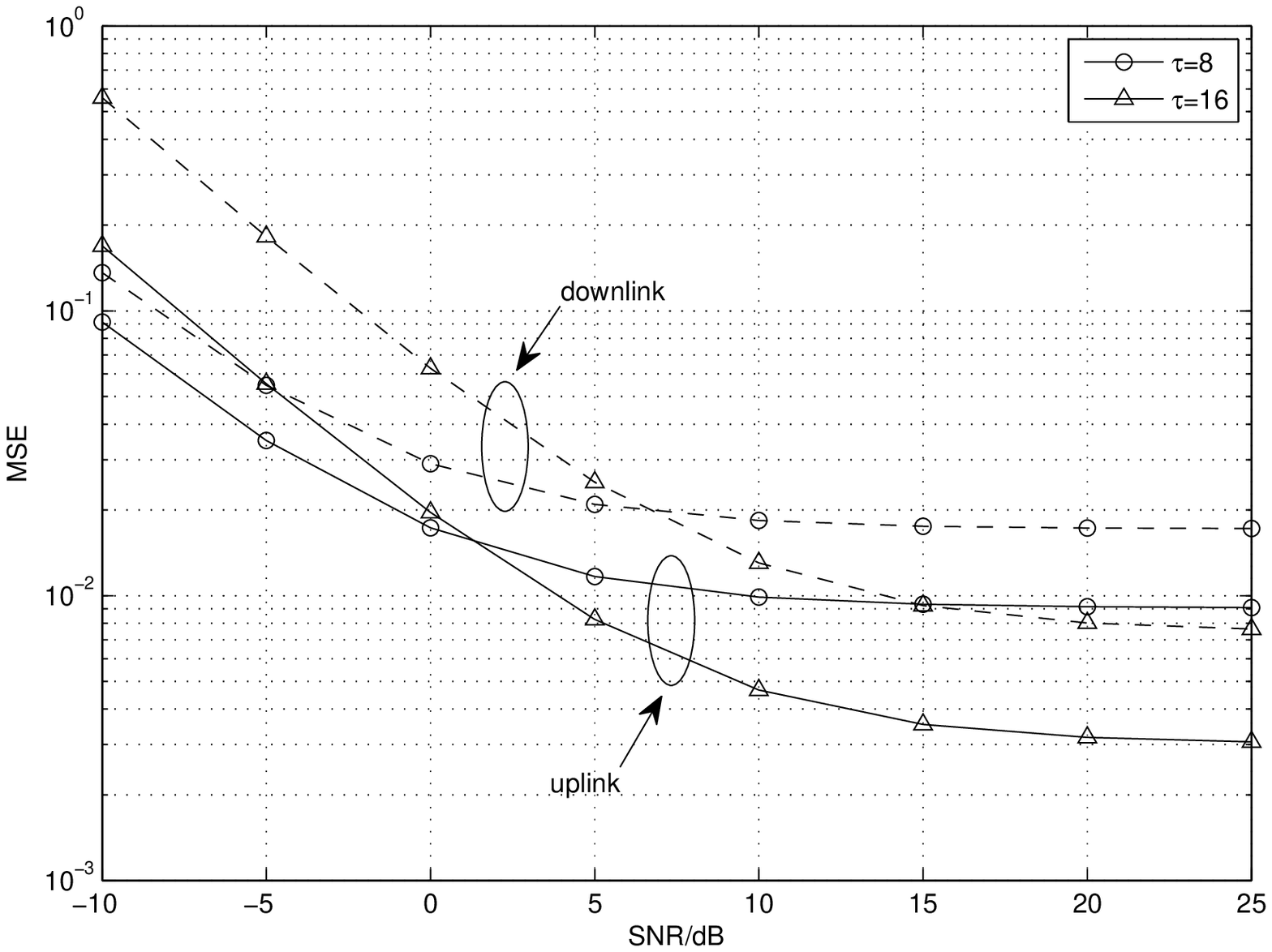}
\caption{ Comparison of uplink/downlink MSE performances with $L=32$ and $\tau=8,16$, respectively.
\label{fig:ULDLMSE_tau}}
\end{figure}

Fig. \ref{fig:ULDLMSE_tau}  displays
the uplink/downlink MSEs as a function of SNR for $\tau=8,16$ respectively with $L=32$.
The total power for both uplink and downlink training is constrained to $L\rho$ for all users at a given SNR $\rho$.
It is seen that as $\tau$ increases, the error floors of uplink and downlink MSEs will decline as expected.
Interestingly, for downlink training, larger $\tau$ will perform better than the smaller $\tau$ at higher SNR but will perform worse at lower SNR.
The reasons might be inferred from \eqref{equ:dlmse} where the three error components  are all closely related to the value of $\tau=|\mathcal{B}_k^{\tu{ro}}|$.
When SNR decreases, the noise component will dominate \eqref{equ:dlmse}
and thus larger $\tau$ will bring more error.
While for higher SNR, the truncation error  will dominate \eqref{equ:dlmse}, and thus smaller $\tau$ will bring more error.
Similar phenomena can also be observed for  uplink training, which can be explained in the same manner over the equation  \eqref{equ:uplinktrainingerror}.

\begin{figure}[t]
\centering
\includegraphics[width=100mm]{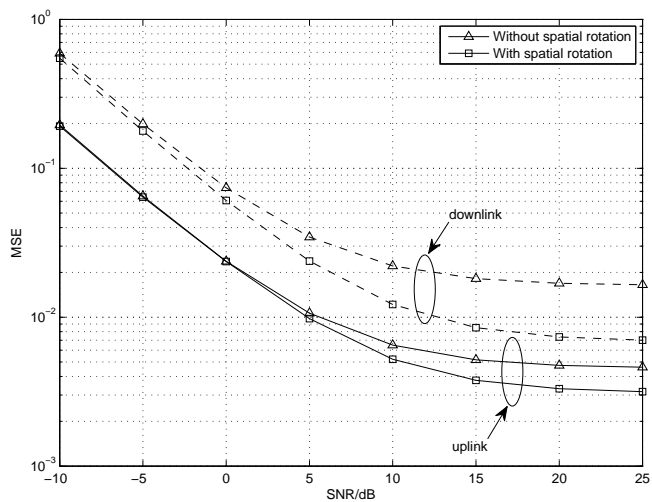}
\caption{ Comparison of uplink/downlink MSE with and without spatial rotation,
where $\tau=16$ and $L=32$.
\label{fig:DLMSE_phaserotation}}
\end{figure}

As spatial rotation  \eqref{equ:phaserotationobjective} is a key technique of our newly proposed strategy, it is then of interest to see
how spatial rotation helps improve the performance. We display the channel estimation for both uplink and downlink with and without spatial rotation in Fig. \ref{fig:DLMSE_phaserotation}.
When the spatial rotation is not adopted, the shift parameter
$\phi_k$ in \eqref{equ:phaserotationobjective} is set as $\phi_k=0$ and then the  channel estimation
procedures remain the same.
It is clearly seen from Fig. \ref{fig:DLMSE_phaserotation} that the spatial rotation will improve the channel
estimation accuracy for both uplink and downlink mainly at high SNRs, i.e., when error floor happens. As explained in
\emph{Property \ref{property:rotation}}, spatial rotation  concentrates more channel
power on fewer DFT points, reduces the power leakage outside $\mathcal{B}_k^{\tu{ro}}$, and thus decreases the truncation
error of  SBEM. Hence, the performance gain by spatial rotation is larger at high SNRs when the truncation error is dominant,
while it is not so obvious at the lower SNRs when the noise dominates.

\begin{figure}[t]
\centering
\includegraphics[width=100mm]{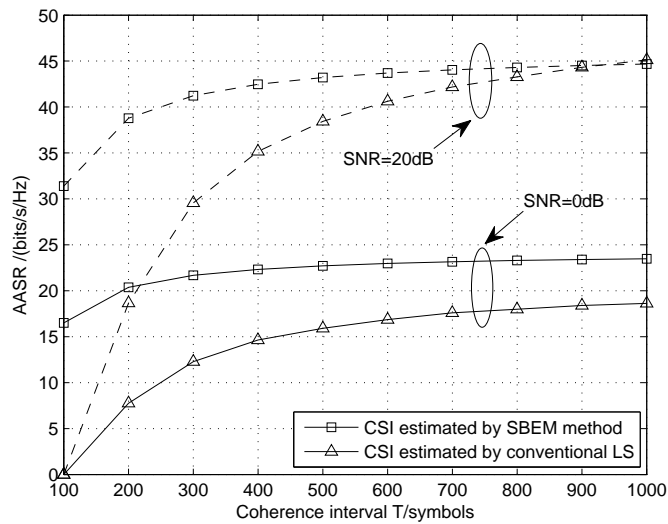}
\caption{The average achievable sum rate (AASR) of the proposed SBEM and conventional LS as a function of coherence interval $T$ with $L=32$.
\label{fig:SE}}
\end{figure}
\begin{figure}[t]
\centering
\includegraphics[width=100mm]{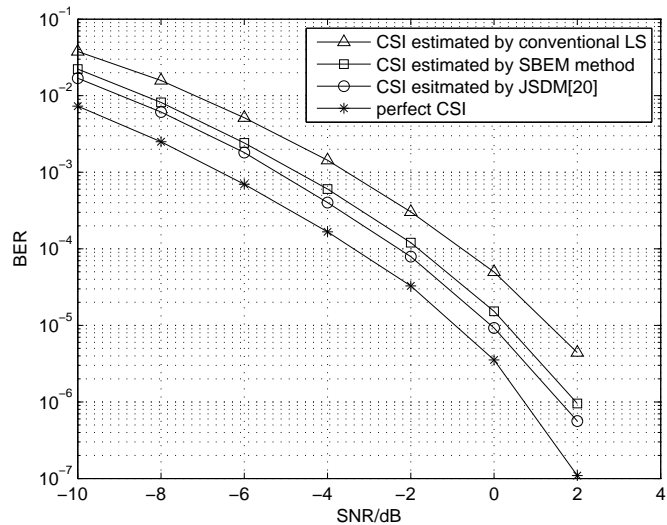}
\caption{Comparison of BER performances with perfect CSI, CSI estimated by JSDM\cite{Caire}, CSI estimated by SBEM, and CSI estimated by conventional
LS method, respectively, where $\tau=16$ and $L=32$.
\label{fig:BER}}
\end{figure}

Fig. \ref{fig:SE} illustrates the average achievable sum rate (AASR) for the downlink data transmission, defined as
\begin{align}\label{equ:SE}
  \textup{AASR}= \left(1-\frac{T_{\textup{pilot}}}{T}\right)\sum_{g=1}^{G^{\tu{dd}}}R(\mathcal{U}_g^{\tu{dd}}|\mathcal{P})/G^{\tu{dd}},
\end{align}
where $T_{\textup{pilot}}$ denotes the length of pilot used for training. We take $T_{\textup{pilot}}=G^{\textup{dt}}L=32$ for the proposed SBEM and  $T_{\textup{pilot}}=M=128$ for the conventional LS.
To make the comparison fair, the overall training power and the overall data power within the coherent time $T$  are
set as the same for each method.
The downlink training procedures for both methods are similar to Fig. \ref{fig:DLMSEcompare} with $\tau=16$ and $L=32$.
Users are scheduled by Algorithm 1.
It can be seen from Fig. \ref{fig:SE} that
the AASR of the proposed SBEM is much higher than that from conventional LS when $T$ is relatively small or when SNR is relatively low. When
$T$ becomes large, the training length of LS is small compared to $T$ and then the AASR from conventional LS
will approach that from the the proposed SBEM. Nevertheless, the main disadvantage of the conventional LS method mainly lies in
the demand of $L=M$ orthogonal training sequences and the high computational complexity.

Next, we show the bit error rate (BER) performance under QPSK
modulation for the downlink data transmission in Fig. \ref{fig:BER}. Four kinds of CSI are compared,
i.e., perfect CSI, CSI from the proposed SBEM method, CSI from the conventional LS method, and CSI from
the joint spatial division multiplexing (JSDM) \cite{Caire}. To keep the comparison fair, the overall training power is set as the same for each method.
It is seen that the BER achieved by JSDM is better than the proposed method by about 0.5 dB due to its exploitation of
real channel covariances matrices. However, JSDM needs $M\times M$ downlink channel covariance matrices of all users and needs to perform  EVD for  all these channel covariance matrices to obtain
the exact basis vectors of the channels.
On the other side, the proposed SBEM utilizes constant Fourier basis vectors for all different users, which eliminates the demand for the downlink channel covariance matrix, and the selection of the spatial signatures can be achieved by efficient FFT. Moreover,
both the low-rank methods, SBEM and JSDM, perform better than the LS method at the low SNR region due to the inclusion of less noise power during channel estimation.

\begin{figure}[t]
\centering
\includegraphics[width=100mm]{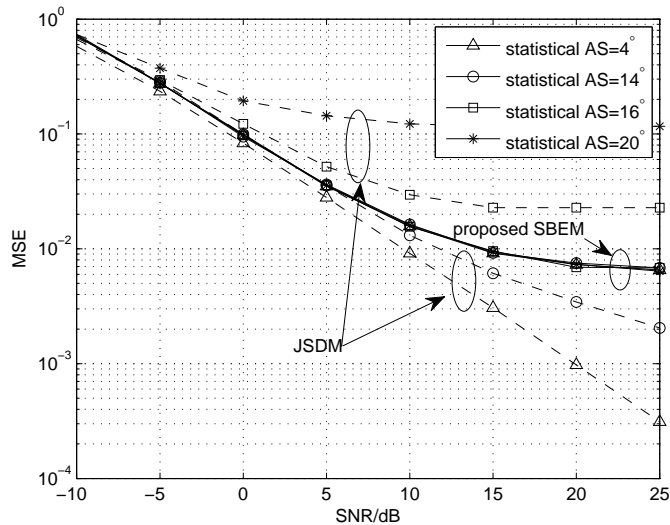}
\caption{The DL MSE performance comparison of SBEM and JSDM \cite{Caire} under the case of user mobility, where
$\tau=16,L=32$.
\label{fig:DLMSEcairemotion}}
\vspace{-2em}
\end{figure}

Lastly, we consider the situation with user mobility.
Fig. \ref{fig:DLMSEcairemotion} compares the DL MSE performances of the proposed SBEM
and the covariance-based JSDM \cite{Caire},
where the instantaneous AS is set as $4^\circ$ at each time and the statistical AS\footnote{The statistical AS is obtained
as user moves around during the measurement period.}
is set as $4^\circ,14^\circ,16^\circ,20^\circ$, respectively.
It can be seen that as the statistical AS increases, the MSE performances of JSDM deteriorate obviously while
the MSE curves from SBEM are not affected so much. The reason lies in
that as the statistical AS increases, channel covariances will cover too broad AS
and thus are not accurate for the instantaneous channel estimation. Instead, SBEM does not
rely on the statistics and thus is more suitable for the mobile user cases.
 
\section{Conclusions}\label{sec:conclusions}

In this paper, we investigated the uplink/downlink training and user scheduling
for multiuser massive MIMO systems. We exploited the physical characteristics
of ULA and proposed a simple DFT-based SBEM to represent the channel vectors
with reduced parameter dimensions. It is shown that the basis vectors for different channels
formulate the spatial beams towards the users, and thus users
could be spatially separated during both the training and data transmission. The conventional headache of pilot contamination is then
immediately relieved. Meanwhile, the uplink spatial signatures could also be used for downlink training based on the
angle reciprocity of electromagnetic propagation, making the proposed SBEM applicable for both TDD and FDD massive MIMO systems.
Channel estimation algorithm for both uplink and downlink were carried out
with very few  training,  and the amount of the feedback could be significantly reduced.
To enhance the spectral efficiency, we also proposed a user scheduling algorithm during the
data transmission period, where the spatial signatures of users were
exploited again.  Compared to existing low-rank models, SBEM provides a new simple way to
determine the spatial signature information for each user without need of
channel statistics and can be efficiently deployed by the FFT operations. Various
numerical results were provided to demonstrate the effectiveness and the superiority of the proposed method.


\linespread{1.2}

\end{document}